\documentclass[prb,twocolumn,amsmath,amssymb,floatfix,tighten,bibtex]{revtex4-1}
\usepackage{graphicx}
\usepackage{bm}
\usepackage[usenames]{color}
\usepackage{epsfig}
\usepackage{subfigure}
\def\Omegas{{\bf\Omega}}

\def\rr{\right}
\def\l{\left}
\def\ba{\begin{array}}
\def\ea{\end{array}}
\def\tal{\tilde{\alpha}}

\newcommand{\sbra}[1]{\langle #1 |}
\newcommand{\sket}[1]{|#1\rangle}

\newcommand{\sex}[1]{\langle #1 \rangle}

\newcommand{\pa}[2]{\frac{\partial #1}{\partial #2}}

\newcommand{\leftexp}[2]{{\vphantom{#2}}^{#1}{#2}}
\marginparwidth=5cm \newcommand{\be}{\begin{equation}} \newcommand{\ee}{\end{equation}} 

\begin{document}

\title {Adiabatic manipulations of Majorana fermions in a three-dimensional network of quantum wires}

\author{Bertrand I.\ Halperin}
\affiliation{Department of Physics, Harvard University, Cambridge, MA 02138, USA}

\author{Yuval Oreg}
\affiliation{Department of Condensed Matter Physics, Weizmann Institute of Science, Rehovot, 76100, Israel}

\author{Ady Stern}
\affiliation{Department of Condensed Matter Physics, Weizmann Institute of Science, Rehovot, 76100, Israel}

\author{Gil Refael}
\affiliation{Department of
Physics, California Institute of Technology, Pasadena, California 91125, USA}

\author{Jason Alicea}
\affiliation{Department of Physics and Astronomy, University of California, Irvine, CA 92697, USA}

\author{Felix von Oppen}
\affiliation{Dahlem Center for Complex Quantum Systems and Fachbereich Physik, Freie Universit\"at Berlin, 14195 Berlin, Germany}

\begin{abstract}

It has been proposed that localized zero-energy Majorana states can be realized in a two-dimensional network of quasi-one-dimensional semiconductor wires that are proximity-coupled to a bulk superconductor.  The wires should have strong spin-orbit coupling with appropriate symmetry, and their electrons should be partially polarized by a strong Zeeman field.  Then, if the Fermi level is in an appropriate range, the wire can be in a topological superconducting phase, with  Majorana states that occur at wire ends and at $Y$ junctions, where three  topological superconductor segments may be joined.
Here we generalize these ideas to consider a three-dimensional network.   The positions  of Majorana states can be manipulated, and their non-Abelian properties made visible, by using external gates to selectively deplete portions of the network, or by physically connecting and redividing  wire segments.  Majorana states can also be manipulated by reorientations of the Zeeman field on a wire segment, by physically rotating the  wire about almost any axis, or by evolution of the phase of the order parameter in the proximity-coupled superconductor.  We show how to keep track of sign changes in the zero-energy Hilbert space during adiabatic manipulations  by monitoring the evolution of each Majorana state separately, rather than keeping track of the braiding of all possible pairs.  This has conceptual advantages in the case of a three-dimensional network, and may be computationally useful even in two dimensions, if large numbers of Majorana sites are involved.

%
%
%
%
%
%
%
%
%
%

\end{abstract}
 \maketitle
\newpage
\section{Introduction}

The last several years have witnessed extensive research into the possibility of realizing zero-energy Majorana-fermion states ("Majoranas") in condensed matter systems. Driven by the interest in non-abelian states of matter and by possible implementations for topological quantum computation~\cite{Kitaev-TQC,Nayak-TQC}, several physical systems have been arenas for the search for Majorana fermions. Initially, these systems were primarily  two dimensional, such as the $\nu=5/2$ fractional quantum Hall state, in which Majorana fermions are believed to be attached to quarter-charged quasi-particles and quasi-holes in the Pfaffian or anti-Pfaffian states~\cite{MooreRead,ReadGreen}; two-dimensional spin polarized p-wave superconductors~\cite{Volovik-pwave,Beenakker-ES}; surface states of three-dimensional topological insulators coupled to superconductors~\cite{FuKane3d}; vortex cores in hybrid systems made of semiconductors with strong spin-orbit interaction in proximity-coupling to superconductors~\cite{Sau2DEG,Alicea2DEG}; and more. While most of the research has been theoretical, interferometry experiments (according to suggestions such as~Refs.\ \onlinecite{ShtengelIF,SternIF}) have yielded data that may be interpreted as resulting from the presence of such excitations~\cite{Kang11,Kang11a,Willett}.

One dimensional systems have been predicted to host Majorana modes as well. Ideas of end-point Majoranas in spin polarized p-wave superconducting wires were already discussed by Kitaev~\cite{Kitaev01} nearly a decade ago using a toy tight binding model for a spin-polarized p-wave superconductor. Possible realizations in semiconductor quantum wires with strong spin-orbit coupling,  such as  InAs nanowires, in the presence of a strong Zeeman field and proximity-coupling  to a bulk s-wave superconductor,  and their manipulation by means of external gates potential, were suggested in Refs.~\onlinecite{Oreg10,Lutchyn10a,Hassler10,Potter10}. One-dimensional systems of the required form might also be realized in edge states formed at the surface of a  topological insulator coupled  to a superconductor  and a ferromagnet~\cite{FuKane-edge,Fu09}.

The presence of localized zero energy Majorana fermions states separated from each other by distances large compared to the coherence length makes the ground state of the system degenerate, in all dimensions. It is well known that in two-dimensional systems braiding Majorana modes around each other produces unitary transformations within the ground state subspace which generally do not commute; hence the statistics is non-Abelian~\cite{Ivanov,SternvonOppen}. More recently, it has been shown that similar manipulations leading to non-Abelian statistics are possible in two-dimensional networks synthesized from wires joined together with $Y$ or $T$ junctions~\cite{Alicea10,Clarke10,Sau11,VanHeck11}.

\begin{figure}[h]
\begin{center}
 \includegraphics*[width=\columnwidth]{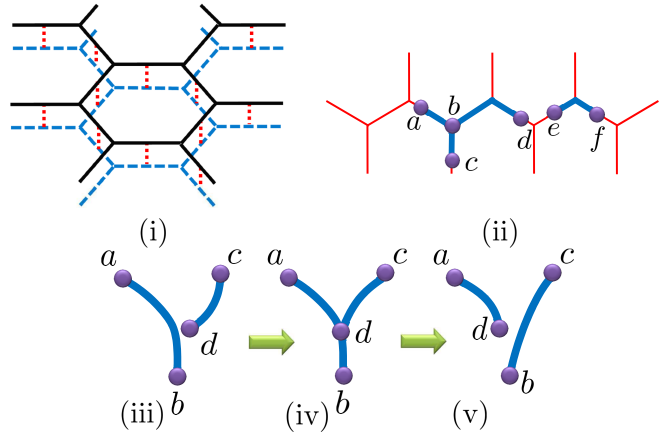}
\end{center}
\vspace{-.5cm}
\caption{\label{fg:wires} (Color online.) Schemes for manipulating Majorana sites. Panel (i) shows two layers  of a three-dimensional network of wires connected by $Y$ junctions.  Solid lines indicate wires in the first layer, dashed lines indicate wires in the second layer, and dotted lines indicate wires connecting the layers.  All wires are tunnel-coupled to a bulk s-wave superconductor with a definite superconducting phase.  Gates (not shown) allow one to control the electron density in individual wire segments so that they can be  either in a topological superconducting (TS) state or are depleted of electrons.
Panel (ii) shows a portion of one layer with depleted regions (thin lines) and occupied regions in the TS state (thick lines).  Labels $a$-$f$ show  Majorana locations, which occur at the end of a TS segment or at a junction where three TS segments are joined.  All wire segments must
be very long compared to the coherence length in the superconducting segments in order for the interactions between Majorana variables to be neglected.      Panels (iii)- (v) show an alternate scheme, where flexible TS wires may be joined together with a $Y$ junction, and subsequently redivided in a different way. Labels $a$-$d$ indicate Majorana positions.  Panel (iii) shows an initial state where $a$ is on the same wire as $b$, while $c$ is joined to $d$.  Panel (v) shows a final state where $a$ is joined to $d$ and $b$ is joined to $c$.
}
\end{figure}

In the present work, we examine a wider set of networks constructed from  one-dimensional wires and their junctions.  The wires and the networks need not be confined to a plane. The wires may be bent out of the plane and may overpass each other, forming a three-dimensional network that is topologically distinct from a two-dimensional system.  Fig.~\ref{fg:wires} illustrates a bilayer network of this type, though there is no barrier to forming an infinite stack of layers which is genuinely three-dimensional.  The network is fixed in space and with permanent junctions, but we imagine that the electron densities in different parts of the network can be  varied on a fine spatial scale by means of external gates, so that segments can be tuned at will between regions of topological superconductor (TS) and regions which are depleted of carriers and are, consequently, not in a topological superconducting state. For alternative ways to manipulate Majorana fermions in quantum wires, see Ref.\ \onlinecite{Romito11,VanHeck11}.
Another possibility, which is conceptually interesting but may be more difficult to realize in practice, is suggested in the bottom half of  Fig.~\ref{fg:wires}.  Here we envision  a collection of flexible TS wire segments, which may be bent or tilted out of the plane at will,  and which can be joined by junctions and redivided in different ways.  For example, we imagine a process, illustrated in the figure, where one end of wire 1 is attached to the middle of wire 2, and then the joint is broken in a way that one half of wire 2 remains attached to the end of wire 1,  while the other half becomes a disconnected segment. In either case, we require that the various wire segments are all proximity-coupled to a single bulk superconductor, so that  any differences in the phase of the induced  superconducting order parameter between various parts of the network are strictly controlled. Further we require that the manipulations are done in such a way that the topological superconducting gap does not close.

One of  the unique features of a network of one-dimensional wires is that for an isolated wire segment, which is tunnel-coupled to the bulk superconductor but otherwise separated from the rest of the system, the parity of the electron number in the segment is conserved during adiabatic  manipulations, and within the low-energy Hilbert space, the electron number parity can be related to the product of Majorana operators at the two ends of the segment. (The precise relation depends on sign-conventions, which must be carefully defined, as we discuss below.) If two occupied wire segments are adiabatically connected through a $Y$ junction, and subsequently redivided, the total number parity will be conserved, but the parities of the individual wire segments may change.  We show how the relations between  electron parity and Majorana operators can be exploited
to predict the results of manipulations in which a Majorana site is moved through a $Y$ junction connecting several wires.

We  also consider a variety of other manipulations that can alter the state in the zero-energy Hilbert space  spanned by the Majorana operators.
For the networks of semiconductor quantum wires discussed here, there are  possible ways to manipulate the Majorana fermion state that do not have direct counterparts in a continuum  two-dimensional system.  In addition to the possibility of moving the Majorana locations in three dimensions, we find that the state may be manipulated by a rotation of the Rashba (spin-orbit-controlling) electric field (together with the magnetic field) around an axis parallel to the wire, or a rotation of the Zeeman magnetic field, provided that the Zeeman field never points along the axis
perpendicular to the plane containing the Rashba field and the wire direction. (Technically, the magnetic field should not come too close to this axis if one is to preserve the gap.)

  As in a two-dimensional network, the Majorana state may also be manipulated by means of  controlled rotations of the superconducting phase difference between different parts of the network, which may be implemented  by moving vortices through the bulk superconductor in regions between the superconducting nanowire segments.    We also consider processes where field orientations are twisted along the length of a wire segment, or the wire itself is tied in a knot.

We note that the behavior of Majoranas passed through a $Y$ junction and the behavior under rotations of the phase of the superconducting order parameter have previously been studied in
 Ref.~\cite{Alicea10} mainly using Kitaev's tight binding toy model~\cite{Kitaev01}. The latter model is topologically equivalent to the semiconductor-superconductor wire and a direct mapping exists for very strong Zeeman field~\cite{Alicea10}.  Exchange processes have been studied in details by Clarke, Sau and Tewari~\cite{Clarke10, Sau11}.  Our findings are in agreement with  the previous works, although our methods are somewhat different. Our detailed study of the combination of spin-orbit interaction with the Zeeman field allow us to analyze new types of manipulations, that are not possible in the context of Kitaev's chain model. Furthermore, our detailed study of the relation between the Majorana modes and the parity of the number of electrons within each wire leads us to conclusions regarding the physical observables associated with a series of braidings. The possibility of manipulating Majoranas in three dimensions has previously been discussed in a general way by Teo and Kane~\cite{Teo10} and by Freedman et al. \cite{Freedman11a,Freedman11}, but not in the specific context of a wire network.

 We imagine initializing the system in a state where each wire has an even number of particles. This may be carried out if the wires are populated from full depletion, by Cooper pair tunneling from a superconductor. As electrons will enter the wire only in pairs, the resulting occupied section will be in a state of even parity. Following the initialization we imagine a series of adiabatic manipulations, at the end of which the parity of the number of particles in each wire is measured. Evidently, the parity state does not depend on the exact location of the  Majoranas at the end points of the wire and hence the manipulations are not required to correspond to closed trajectories of any kind. We find that the set of states to which the system may be brought by topological manipulations is rather limited. After such manipulations, each wire may end up either in a pure state of even or odd parity, or in a mixture of even and odd parities, at equal weights. No other states may be constructed in a topological manner.

Our analysis involves also a technical part. We show that the evolution of the system may be tracked by two equivalent methods. One method is based on a single-valued parametrization of the Majorana operators in terms of the time-dependent parameters of the system's Hamiltonian. We find that a single-valued parametrization of the Majoranas necessarily involves discontinuities at arbitrarily chosen cuts. While these discontinuities are obviously absent from the physically observable quantities, they are present in the relation between the parity operators and the Majorana operators. The second method uses a parametrization of the Majorana operators that is multi-valued with respect to the parameters of the Hamiltonian. In this parametrization there are no discontinuities, but its being multi-valued makes it dependent on the history of the system. We show that the two methods are equivalent, but most of our analysis is carried out using the first one.

The structure of the article is as follows. In Section \ref{se:Models}, we define the models to be considered,  discuss the relation of  Majorana operators to the creation and annihilation operators for electrons, and define the sign conventions to be used in the paper. Specifically, in Subsection
 \ref{se:ModelsN} we discuss the general properties of the Majoranas in a wire with $N$ spin resolved electron modes below the Fermi energy. In Subsection~\ref{se:ModelsN1} we  consider  the $N=1$ case of  a spin -polarized electron system coupled to a $p$-wave superconductor, and we discuss explicit solutions for the Majoranas at the end of a wire in that case.
In Subsection~\ref{se:ModelsN2},  we analyze a quantum wire of electrons subjected to spin-orbit coupling  and a Zeeman field, coupled by proximity to an $s$-wave superconductor, which corresponds to $N=2$ by our conventions. At strong magnetic field a projection of the $N=2$ wire onto the low energy band reduces to the $N=1$ model.

In Section~\ref{se:onewire} we explain the methods with which we analyze manipulations involving Majorana  modes at the ends of a single wire. We first consider the general principles that determine the adiabatic time evolution of Majorana fermion operators. In particular we show that any observable which is a product of zero energy Majorana operators is independent of time except at isolated instants when it may be multiplied by $-1$. We then analyze carefully the single-valued vs multi-valued representation of the Majorana end modes, and relate the Majorana operators to physical observables, in particular the parity of the number of electrons in each wire. Subsection \ref{examples subsection} gives examples for manipulations of one wire and their resulting transformations.

In Section~\ref{se:BasicY} we discuss explicitly the possible sign changes that result when a  Majorana is transferred through  a $Y$ junction, from one wire to another, as well as their physical significance.    In Section \ref{se:MajornaExchange} we use these results to study the sign changes and the physical consequences that occur when the locations of two Majoranas  are exchanged by manipulation through a $Y$ junction. We examine both  the case where the two Majoranas belong to different topological superconductor segments and the case where they are at opposite ends of a single topological segment.  Results are summarized in Section~\ref{se:Summary}.

\section{Explicit Models of Topological superconducting wires}
\label{se:Models}
In this section we will review some explicit models for $N$-mode wires that can form a TS phase. We will discuss the structure of the zero energy state Majorana solutions at the end of the wires in several examples.
\subsection{N-mode quantum wires proximity-coupled to a bulk superconductor}
\label{se:ModelsN}
Majorana fermion excitations are known to appear in quantum wires which are proximity coupled to superconductors~\cite{Kitaev01,Oreg10,Lutchyn10a,Potter10}. Such systems are generally described by pairing Hamiltonians,
\begin{eqnarray}
H &=& \int dw \left[\Psi_\alpha^\dagger (w)h_0^{\alpha\beta}\Psi_\beta (w) \right.\nonumber \\
  &+& \left.\Psi_\alpha^\dagger (w)\Delta_{\alpha\beta}\Psi_\beta^\dagger (w)+h.c.\right].
\label{ham}
\end{eqnarray}
These Hamiltonians are quadratic in the electronic creation and annihilation operators $\Psi^\dagger_\alpha(w)$ and $\Psi_\alpha(w)$, where $\alpha = 1,\ldots , N$ is a spin and/or channel index and $w$ denotes the coordinate along the wire. (We will have occasion to consider wire orientations along arbitrary directions in space and reserve the notation $x,y,z$ for the coordinates along the fixed laboratory coordinate axes.) The wire in the normal state is described by the single-particle Hamiltonian $h_0^{\alpha\beta}$ which includes the kinetic term, a scalar potential, as well as the spin-orbit and Zeeman couplings. The superconducting proximity effect is accounted for by the pair potential $\Delta_{\alpha\beta}$ which may depend on the momentum $p_w$. While the single-particle Hamiltonian $h_0$ is hermitian, $h_0^\dagger = h_0$, the Pauli principle demands that the pair potential is antisymmetric, $\Delta=-\Delta^T$.

The excitation spectrum of the Hamiltonian (\ref{ham}) as well as the associated quasiparticle (Bogolyubov) operators $\gamma_E$
can be obtained with the ansatz
\begin{equation}
\gamma_E = \int dw [u_\alpha^*(w) \Psi_\alpha(w) + v_\alpha^*(w) \Psi_\alpha^\dagger(w)]
\end{equation}
where the index $\alpha$ is summed over and the condition that $\gamma_E$ be an eigenoperator of $H$, $[H,\gamma_E] = - E \gamma_E$. This yields the Bogolyubov-de-Gennes equation
${\cal H}\psi=E\psi$ with the $2N\times 2N$ Bogolyubov-de-Gennes Hamiltonian
\begin{equation}
{\cal H}=\left (\begin{array}{cc} h_0 & \Delta \\ \Delta^\dagger & -h_0^T \end{array} \right)\label{hbdg}
\end{equation}
and $\psi=[{\bf u}(w),{\bf v}(w)]$. Here, we collected the components $u_\alpha$ and $v_\alpha$ into spinors ${\bf u}$ and ${\bf v}$.

As usual, the spectrum of the Bogolyubov-de-Gennes Hamiltonian is particle-hole symmetric. Indeed, ${\cal H}$ obeys the relation ${\cal H} = -\tau_x H^T \tau_x$, where $\tau_x$ denotes a Pauli matrix in particle-hole space. Thus, if $|\psi\rangle$ is an eigenvector with energy $E$, then $K\tau_x|\psi\rangle$ is an eigenstate with energy $(-E)$, where $K$ denotes the complex-conjugation operator. Under certain conditions, the Bogolyubov-de-Gennes (BdG) equations possess zero-energy solutions, which correspond to Majorana fermions. These solutions satisfy (with an appropriate choice of phase) $|\psi\rangle = K\tau_x |\psi\rangle$, which implies that
 ${\bf u}(w) = {\bf v}^*(w)$. Introducing the notation $g_\alpha(w) = u^*_\alpha(w)$ for the Majorana solutions, the associated Bogolyubov operator becomes
\begin{equation}
\gamma = \int dw \left[g_\alpha(w)\Psi_\alpha(w) + g^*_\alpha(w)\Psi_\alpha^\dagger(w)\right] ,
\label{majorana-def}
\end{equation}
which satisfies the Majorana-fermion property $\gamma = \gamma^\dagger$. In principle, a solution for $\gamma$ could be multiplied by an arbitrary phase factor and still commute with the Hamiltonian.  However, the convention that a Majorana operator should be self-adjoint limits the allowable phase factors to $\pm1$.

The Hamiltonian $H$ depends on a set of parameters such as the wire orientation or the direction of the applied magnetic and electric fields. We will denote these parameters collectively as $\Omegas$. In this paper, we will focus on processes in which these parameters vary slowly in time, denoting the initial parameters at time $t=0$ as $\Omegas_0$ and the final parameters at time $t=T$ as $\Omegas_T$. We will analyze the resulting adiabatic time evolution of the Majorana states given by $g(w;\Omegas)$ and of the corresponding Majorana operators $\gamma(\Omegas)$, focusing on two specific models which we describe in the remainder of this section.

\subsection{Spinless fermions}
\label{se:ModelsN1}
We first consider the Hamiltonian (\ref{ham}) for $N=1$, corresponding to spin-polarized electrons, with a pair potential $\Delta =  -i v p_w e^{i(\theta+\alpha)}$ which originates from proximity coupling to a bulk $p_x+ip_y$ superconductor. Here, $\theta$ denotes the overall  phase of the order parameter $\Delta$ in the bulk superconductor,  and $\alpha$ is the angle between the direction of the wire (given by the unit vector ${\hat{\bf w}}$) with the $x$-axis.  (We assume for now that ${\hat{\bf w}}$ lies in the $x-y$ plane, so $\cos\alpha={\hat{\bf w}}\cdot{\hat{\bf x}}$,    $\sin\alpha={\hat{\bf w}}\cdot{\hat{\bf y}}$.) We shall choose a branch cut such that $-\pi<\alpha \le \pi$. The Hamiltonian of the normal state includes only a kinetic term which we model with a quadratic dispersion so that
\begin{widetext}
\begin{equation}
   H = \int dw \left[\Psi^\dagger (w)\l(\frac{p_w^2}{2m}-\mu\rr)\Psi(w)+\Psi^\dagger (w) v (-i p_w) e^{i(\theta+\alpha)} \Psi^\dagger (w)+h.c.\right].
\label{pwave}
\end{equation}
\end{widetext}
This model of a one-dimensional spinless $p$-wave superconductor is referred  to as the continuum version of Kitaev's toy model, with Bogolyubov-de-Gennes Hamiltonian
\begin{equation}
{\cal H}=\left[\frac{p_w^2}{2 m}-\mu \right ]\tau_z-i v p_w e^{i\left(\theta+\alpha\right)} \tau_+ + h.c.;
\label{onewireham}
\end{equation}
 The matrices $\tau_a$ denote Pauli matrices, which operate in particle-hole space.

It turns out \cite{Alicea10} that the continuum Kitaev  model (\ref{pwave}) also arises in the context of semiconductor wires with strong spin-orbit coupling in proximity to conventional $s$-wave superconductors. The latter system, which we discuss in the subsection~\ref{se:ModelsN2}, reduces to the continuum Kitaev model in the limit of a strong Zeeman field. The explicit connection between the models is given in Eq.~(\ref{eq:Blarge}).

Majorana modes occur at the edges of the wire when the chemical potential within the wire is
above the bottom of the electronic band ($\mu>0$ in the case of
quadratic dispersion). Substituting $i p \rightarrow \partial_w$ in Eq.~(\ref{onewireham}) we find that the Majorana mode obeys the differential equation (here and throughout the rest of the paper $\hbar=1$) :
\begin{equation}
\label{eq:gwdiff}
-\frac{1}{2m} g''(w)- \mu g(w) - v e^{i (\theta+\alpha)} {g^*}'(w) =0,
\end{equation}
with the boundary condition $g(w=0)=0$; (we assume here that the wire ends abruptly  near the origin of the coordinate system, and the occupied region has $w > 0$).
For $k_F> mv$, the solution takes the form:
 \begin{eqnarray}
 \gamma&=&\int d w g(w)  \Psi(w) +h.c.;  \nonumber\\
  g(w)&=&  e^{-i([\alpha+\theta]_r  )/2} 2 \sqrt{m v}\sqrt{\frac{k_F^2}{k_F^2-(m v)^2}} e^{-w m v} \nonumber \\
  &\times& \sin\left[w \sqrt{k_F^2-(m v)^2}\right],
 \label{eq:majneq1}
\end{eqnarray}
where $\mu = k_F^2/2m$, and $[x]_r$ is equivalent to $x$ modulo $2\pi$.  This leaves an overall sign ambiguity in the definition of
$\gamma$, which we resolve, in the following, by choosing $[x]_r$ to lie
in the interval
\begin{equation}
\label{eq:modx}
- \pi < [x]_r \leq \pi  .
\end {equation}
Equivalently, we may write
\begin{equation}
\label{eq:modx1}
[x]_r \equiv  [ (x + \pi)\!\!\! \mod 2 \pi ] - \pi ,
\end{equation}
where ``mod"  is defined to give a value in the interval $(0, 2 \pi]$.
With these conventions, we see that for $\theta = \alpha = 0$, the first peak due to the term $\sin\left[w \sqrt{k_F^2-(m v)^2}\right]$ is positive.

\begin{figure}[h] \begin{center}
 \includegraphics*[width=.9\columnwidth]{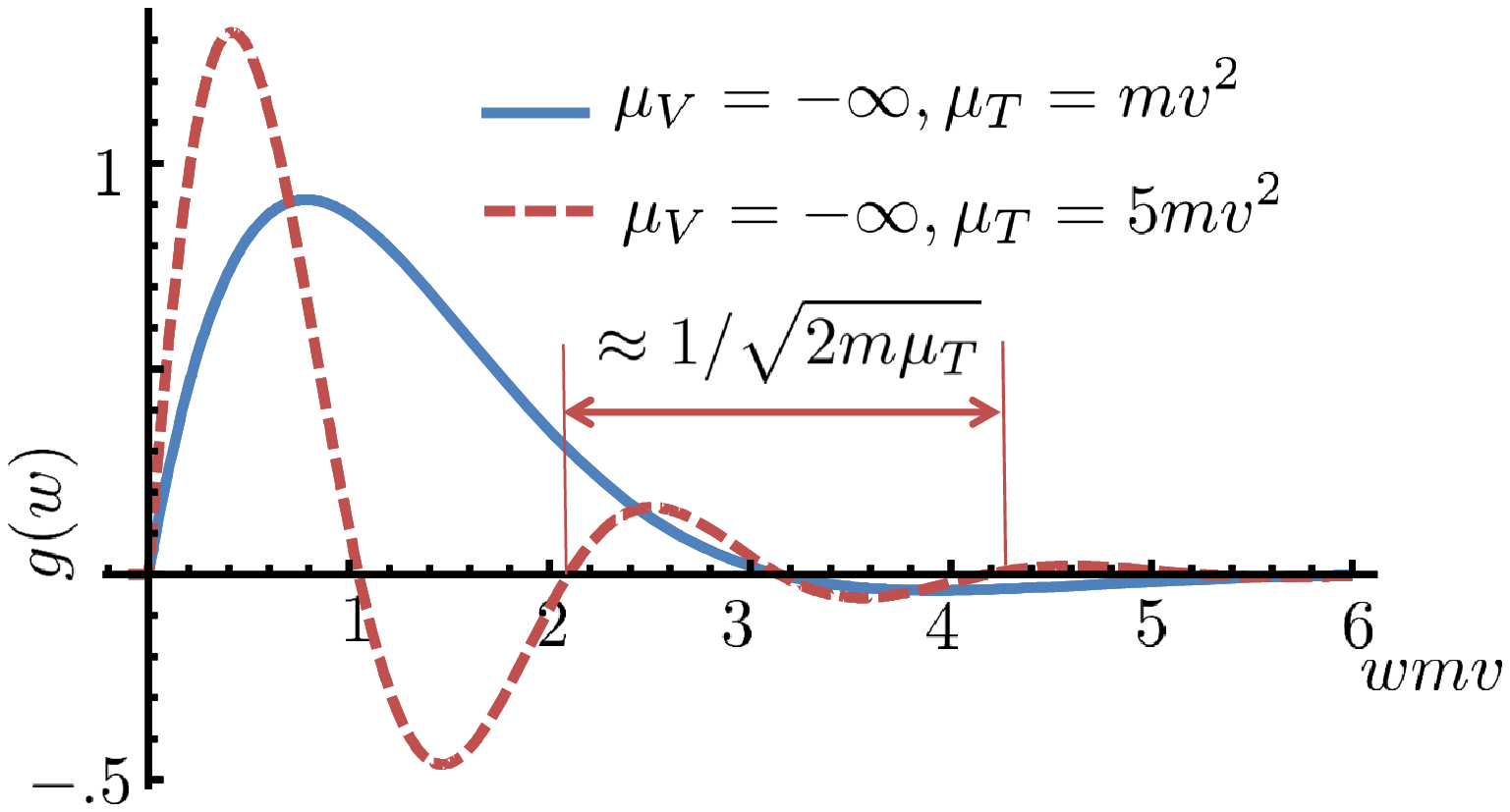}
 \includegraphics*[width=.9\columnwidth]{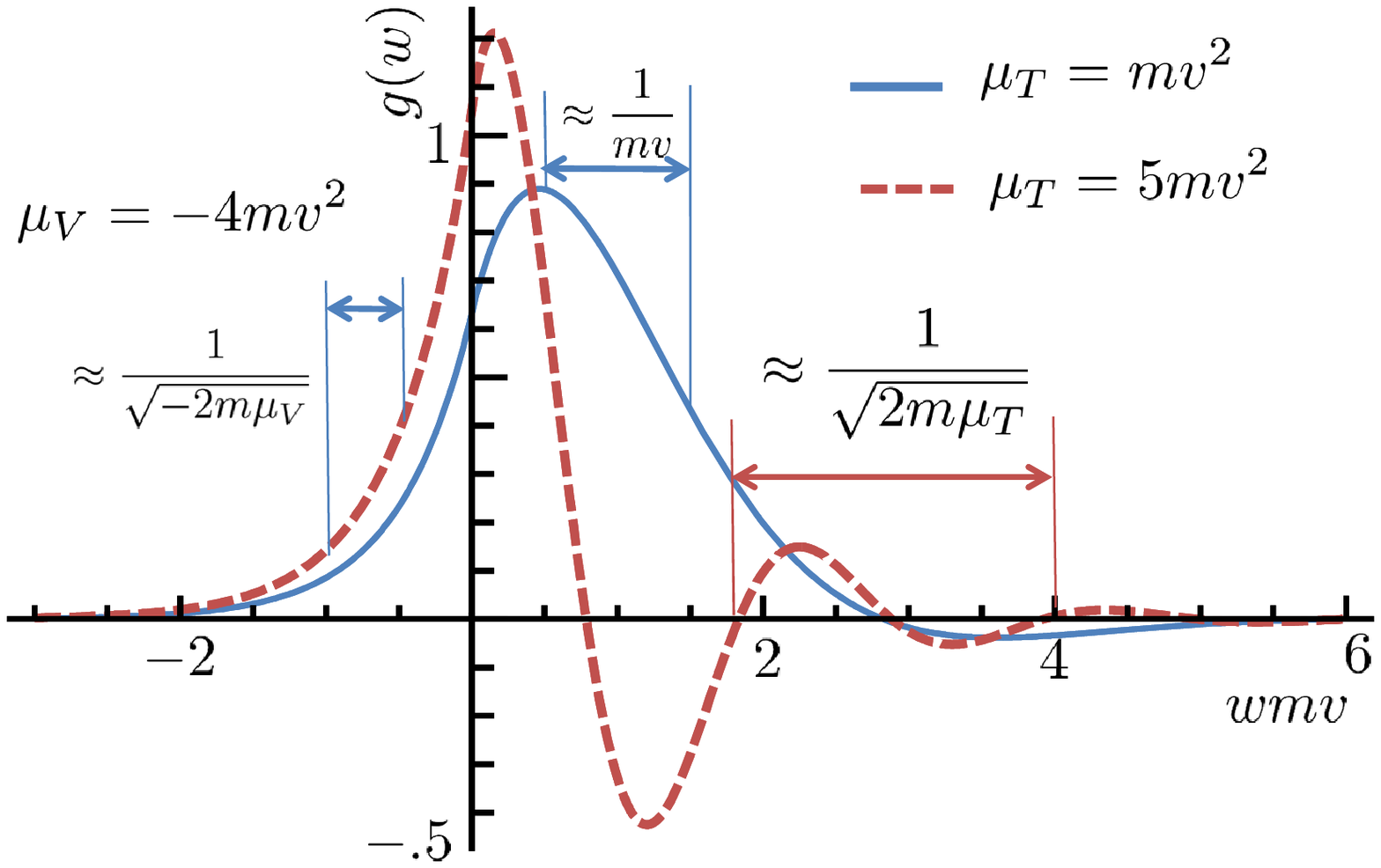}
  \end{center} \vspace{-.5cm} \caption{\label{fg:wireends} Typical
Majorana state at the end of the wire in the N=1 case. In the upper panel the topological state is terminated in a vacuum with infinite potential while in the lower panel it has finite potential $\mu_V$.  The typical length scale for the oscillations and decay rate are depicted in the limit $\mu_T \gg m v^2$ and
$\mu_V \ll - m v^2$. See Eq.~(\ref{onewireham}) for the definition of $v$.} \end{figure}

Majorana states can also be localized in the bulk of the wire by
allowing the local chemical potential  $\mu$ to drop below the bottom of the band. Notice that  since the chemical potential $\mu$ is defined as the difference between an electrochemical potential $\mu_0$ which is independent of position,  and an electrostatic potential $V(w)$ that may depend on position, spatial variations in $\mu$ are produced by variations in $V$.
At the interface where $\mu$ changes sign, a Majorana state will be localized.
Now, we have to solve Eq.~(\ref{eq:gwdiff}) separately for the region with negative chemical potential $\mu_V<0$ and the topological region with $\mu_T>0$ and match the wave function $g$ and its derivative $g'$ at the boundary between them. We assume that  $\alpha+\theta=0$ and  choose units of $m v^2/2$ for energy and $m v$ for momentum. We assume that the  topological phase, with $\mu_T>0$,  occurs for $w>0$ and the non-topological phase (vacuum), with $\mu_V<0$, for $w<0$. In these dimensionless units the equation for the Majorana zero modes becomes:
\begin{equation}
\label{eq:g}
-g''(w)-\mu (w) g(w)-2 g'(w) =0,
\end{equation}
with boundary conditions $g(0^+)=g(0^-)$ and $g'(0^+)=g'(0^-)$.
The full expression for the Majorana state, with a sign choice similar to the one defined above for the abrupt wire end, is
\begin{eqnarray}
g(w) &=& {\cal N}^{-1}
 e^{-w} \left(\cos\kappa_T w  + \frac{\kappa_V}{\kappa_T} \sin \kappa_T w \right) \Theta(w) \nonumber \\
 &+& {\cal N}^{-1}e^{(\kappa_V-1) w} \Theta(-w).
  \label{eq:wflargeB}
 \end{eqnarray}
 Here we assume  $\mu_T>1 \text{ and } \mu_V<0$ and define ${\cal N}^{-1}=2\sqrt{\frac{(1+\kappa_T^2)(\kappa_V-1)}{(\kappa_T^2+\kappa_V^2)(\kappa_V+1)}}$, $\kappa_T^2=\mu_T-1>0$, $\kappa_V^2=1-\mu_V$, and the unit step-function $\Theta(w)$. If $\alpha + \theta \neq 0$, the expression for $g$ should be multiplied by $e^{-i [ \alpha+\theta]_r / 2}$.
 The Majorana wave functions have a simple interpretation: in the ``vacuum" the state decays with inverse decay length
$(\sqrt{1-2 \mu_V /mv^2} -1) m v  \xrightarrow{|\mu_V| \gg m v^2} \sqrt{-2 m \mu_V }$. In the topological phase the state decays with a decay
length $1/(mv)$ set by the induced superconducting gap; on top of this, there are oscillations on the scale of the Fermi momentum in the superconductor
$\kappa_T = m v \sqrt {2\mu_T/ (m v^2)-1} \xrightarrow{\mu_T  \gg m v^2} \sqrt{2 m \mu_T}$. Typical states at the end of the wire are depicted in
Fig.~\ref{fg:wireends}.

If initially the wire is along the $x$-axis then we have two Majoranas -- one with $\alpha=0$ and the other with $\alpha=\pi$. When we rotate the wire by $360^\circ$ the term $e^{-i [ \alpha+\theta]_r / 2}$ will change sign for both Majoranas. However, a rotation of the wire by $180^\circ$ will cause only one of them to cross the cut and hence only one of the Majoranas will be multiplied by a minus sign. We elaborate on this point is subsection \ref{se:AppMaj}.

\subsection{Quantum wires with strong spin-orbit coupling}
\label{se:ModelsN2}

Consider now the case $N=2$, corresponding to spinful electrons in the conduction band of a semiconductor, where the index $\alpha$ enumerates the spin orientations $\uparrow$ and $\downarrow$.  The electrons are subject to a Rashba spin-orbit coupling of strength $u$, originating from an internal electric field that points in direction  $ \hat{\bf e}$ perpendicular to the direction of the wire,  a Zeeman field ${\bf B},$  and a proximity-induced $s$-wave pair potential $\Delta_{\uparrow\downarrow} = |\Delta| e^{i\theta}$.

In order to write the associated  Bogolyubov-de-Gennes Hamiltonian, we find it useful to follow a modified ordering of the Bogolyubov-de-Gennes spinor \cite{Fu08}, namely $\psi(w) = [u_\uparrow(w), u_\downarrow(w), v_\downarrow(w), -v_\uparrow(w)]$. With this ordering, the off-diagonal particle-hole blocks of the Bogolyubov-de-Gennes Hamiltonian become diagonal in spin-space,
\begin{equation}
 {\cal H}=\left[\frac{p_w^2}{2m} - u {\bf p} \cdot {\boldsymbol{ \sigma}}\times \hat{{\bf e}} - \mu\right]\tau_z - {\bf B}\cdot{\boldsymbol \sigma}  + |\Delta|   e^{i\theta} \tau_+ + {\rm h.c.}.
\label{eq:h0}
\end{equation}
Here, $\sigma_i$ denotes Pauli matrices in spin space. In this representation, the particle-hole symmetry of the Bogolyubov-de-Gennes Hamiltonian follows from $\{ {\cal H} , -i\tau_y {\cal T}\} = 0$, where ${\cal T} = i \sigma_y K$ denotes the time-reversal operator and, with an appropriate choice of phase so that $\gamma$ is self-adjoint, a Majorana state will satisfy $|\psi\rangle = -i\tau_y {\cal T} | \psi \rangle$.

Of the parameters $\Omegas$ contained in this Hamiltonian that can be exploited to implement non-trivial transformations within the ground-state manifold, our discussion will focus on the phase $\theta$ of the superconducting pair potential, the direction ${\hat{\bf w}}$ of the wire, the direction of the Zeeman field, and the unit vector ${\hat{\bf e}}$ pointing along the Rashba field.

A necessary condition for the wire to exist in a TS state is that the Zeeman field ${\bf B}$ has a non-zero projection onto the plane formed by $\hat{\bf e}$ and $\hat{\bf w}$. Thus, we may always  define a unit vector $\hat{\bf b}$ pointing in the direction of this projection.  For simplicity, we shall assume in most of our discussions that the component of ${\bf B}$ perpendicular to  the  $(\hat{\bf e},\hat{\bf w})$ plane is zero, so that ${\bf B} = B \hat{\bf b}$.  In the general case, a non-zero perpendicular component  of ${\bf B}$ may be treated as an additional parameter of the Hamiltonian, along with, {\it e.g.},  the magnitude of ${\bf B}$ and the profile of the variation in the chemical potential $\mu$, which will affect  the precise shape of the  wave function associated with the  Majorana operator, but will not be important for the sign conventions and electron parity manipulations we focus on in this paper.

The Hamiltonian (\ref{eq:h0}) can be reduced to simpler models in a variety of limits~\cite{Alicea10,Brouwer11}. In particular, it reduces to the Kitaev model discussed in the previous subsection in the limit of large Zeeman splitting. In this limit, the spin direction at the Fermi points is dominated by the Zeeman field, with only a small correction due to the spin-orbit coupling. The strength of the $s$-wave proximity effect is directly controlled by these small deviations, effectively causing $p$-wave-like correlations in the semiconductor wire.
At large magnetic field $B \gg m u^2$, before the coupling to the superconductor, the two bands of the wire are well separated and only one band crosses the Fermi energy. If we project onto this band, the coupling to the superconductor gives rise to a BdG Hamiltonian of the form~\cite{Alicea10}:
\begin{equation} \label{eq:Blarge}
 {\cal H}= \left( \begin{array}{cc}
  \frac{1}{2 m_{\rm eff}} p^2-\mu_{\rm eff} & -i v_{\rm eff} p e^{i (\theta+\alpha)}  \\
   i v_{\rm eff} p e^{ -i (\theta+\alpha)}  & -\frac{1}{2 m_{\rm eff}} p^2 + \mu_{\rm eff}
\end{array} \right)
\end{equation}
  with $\mu_{\rm eff} = \mu+B, \; v_{\rm eff} = u \Delta/B,\; 1/m^*=1/m(1-mu^2/B)$ and $\cos \alpha = \hat {\bf w} \cdot \hat {\bf x}$. The $2 \times 2$ matrix operates in the particle-hole space (described by $\tau$ matrices). This Hamiltonian represents a spinless $p$-wave superconductor in one dimension. Interestingly the phase of the order parameter depends on the direction $\alpha$ of the wire - this fact will be useful when we will discuss interchange of Majoranas in a $Y$ junction with the wires oriented in specified directions on the surface of the bulk superconductor.


 \begin{figure}[h]
\begin{center}
\includegraphics*[width=.7\columnwidth]{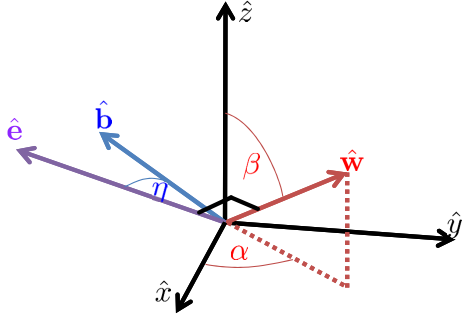}
\end{center}
\vspace{-.5cm}
\caption{\label{fg:coordinates} (Color online)
Orientation  angles defined in the text.  The direction of the wire $\hat{\bf w}$ in the fixed laboratory coordinate system is described by the angles $\alpha$ and $\beta$. The electric field direction $\hat{\bf e}$, which is required to be perpendicular to the wire direction $\hat{\bf w}$, is rotated away from  the $(\hat{\bf z},\hat{\bf w})$ plane by an angle $\delta$, not shown.  Finally,
$\eta$ is the angle between $\hat{\bf e}$ and the unit vector $\hat{\bf b}$, which is the direction of the projection of the Zeeman field  ${\bf B}$ onto the $\hat{\bf e},\hat{\bf w}$ plane. In the text, we examine particularly the example where $\hat{\bf w}$ is in the $x-y$ plane, and $\hat{\bf e} = \hat{\bf b}=\hat{\bf z}$, so that $\beta=\pi/2$, and $\eta = \delta =0$.}
\end{figure}


The $N=2$ case of semiconductor wires with strong spin-orbit coupling, subject to a Zeeman field, allows for a richer set of nontrivial transformations. In addition to rotations and changes of the phase of the order parameter, one can consider variations of the direction of the Zeeman field ${\bf B}$ as well as the spin-orbit field $u \hat{\bf e}$. In addition, we consider rotations of the wire in three dimensions under the assumption that the wire remains proximity coupled to the bulk $s$-wave superconductor at all times.

Away from the strong field limit,  the Majorana solution of the BdG equations for $N=2$ has four components, which we write as $\psi=\frac{1}{\sqrt{2}}(\psi_p,\psi_h)$ where the normalized two-component spinors $\psi_p$ and $\psi_h$ satisfy $\psi_h={\cal T}\psi_p$.
Let us consider a wave function for a Majorana state, labeled by $l$,  at the end of a TS wire segment. Since $\psi_h$ is determined by $\psi_p$, we need only describe the behavior of the latter.

 Let us suppose that the parameters such as the variation in the chemical potential at the end of the wire,  the magnitude of the spin-orbit coupling, and  the magnitude of the Zeeman field $B$,  are fixed, along with the magnitude of the superconducting pairing field $|\Delta|$.  The form of the spinor $\psi^l_p$  will then depend on the orientation of the wire, the orientation of $\hat{\bf e}$ and the orientation of $\hat{\bf b}$.  For definiteness, we choose the wire orientation  $\hat{\bf w}$ to point from the depleted region into the TS region.  Further we may define the variable $w$ to be the distance along the wire taking as the origin the point where the chemical potential passes through zero, with positive $w$ representing the TS side.   Let $\psi_{p0} (w)$ be the spinor wave function in the case where $\hat{\bf w} = \hat{\bf x}$ and $\hat{\bf e}= \hat{\bf b} = \hat{\bf z}$, with real $\Delta>0.$  The function $\psi_{p0}$ is uniquely determined, except for an overall sign, which we choose using a convention that the $\sigma=\uparrow$ component has a positive real part at the point $w$ where the wave function has its maximum amplitude [cf. the explicit forms for the wave function in strong magnetic field limit, which  were discussed in Eqs.~(\ref{eq:wflargeB}) and (\ref{eq:Blarge}) and  illustrated in Fig.~\ref{fg:wireends}].

For any other choice of orientations, we may write
\begin{equation} \label{eq:MF}
 \psi_p^l(w;{\bf{\Omega}}) = {\hat {U}}_l  \psi_{p0} (w)
\end{equation}
where  $\hat{U}_l $ is a $2 \times 2$ unitary matrix. For brevity we will omit the subscript $l$, which enumerates the Majorana operators, throughout the remainder of this section.

Let us first consider the case where $\hat{\bf b}$ is parallel to $\hat{\bf e}$.  (By definition, we have $\hat{\bf e} \perp \hat{\bf w}$.)
 The orientation of the pair $(\hat{\bf w}, \hat{\bf e})$ relative to the original axes $(\hat{\bf x}, \hat{\bf z})$, may be described by an SO(3) matrix $R_{ij}$, whose matrix elements are the Cartesian components of the triad   formed by $\hat{\bf w}, (\hat{\bf e} \times \hat{\bf w})$, and $\hat{\bf e}$.  The matrix can be described by specifying angles $\beta$ and $\alpha$ which are the polar and azimuthal angles of $\hat{\bf w}$ (in the $\hat{\bf x},\hat{\bf y},\hat{\bf z}$ coordinate system), plus an additional rotation angle $\delta$ to describe the orientation of $\hat{\bf e}$ relative to the $(\hat{\bf z}, \hat{\bf w})$ plane (see Fig.~\ref{fg:coordinates}).      Alternatively, one can characterize $R$ as a rotation by a specified angle about a specified axis.  In general, $R$ will have one unit eigenvalue, with a real eigenvector $\hat{\bf d}$, and two other eigenvalues of the form $e^{\pm i \lambda}$.  One can always choose $\lambda$ to lie in the interval $0 \leq \lambda \leq \pi$, and choose the sign of $\hat{\bf d}$ so that $R$ describes a rotation about $\hat{\bf d}$ by a positive angle $\lambda$.  We may then define a unique SU(2) matrix $S$ corresponding to $R$, by
 \be
 S = e ^{- i \lambda \hat{\bf d} \cdot {\boldsymbol\sigma} / 2} \, .
 \ee
Note that if the matrix $R$ is allowed to vary continuously, the matrix $S$ will jump discontinuously whenever the rotation angle $\lambda$ reaches $\pi$, as  $\hat{\bf d}$ will change sign at that point.

Now let us introduce the superconducting phase angle $\theta$,
   $\Delta = |\Delta| e^{i \theta}$ . Then the matrix $\hat{U}_j$ describing the Majorana wave function may be written
\begin{equation}
\hat{U} =  e^{i \theta / 2} S  .
\end{equation}
Again, we are faced with a choice of overall sign, which we must choose by some convention.  We shall do this by defining the angle $\theta$ to be in the range $- \pi < \theta \leq \pi.$

As an example, suppose that  $\hat{\bf e}$ is the z-direction.  Then we must have  $\beta=\pi/2$  and $\hat{\bf w} = (\cos \alpha, \sin \alpha , 0)$. Then
\begin{equation}
S = e^{-i \sigma_z \alpha/ 2}
\end{equation}
Clearly, $\hat{U}$ will have a jump in sign if $\alpha$ crosses the cut at $\alpha= \pm \pi$.   Similarly it will have a jump if   $\theta$ crosses the value $\pm \pi$. We may contrast this with the sign convention
we employed for the $N=1$ wire. In that case, we had only a single cut, so the jump in sign occurs when $[\theta + \alpha]_r$ reaches $\pm \pi$.

 Let us now consider the situation where $\hat{\bf b}$ is not aligned with $\hat{\bf e}$.
  Since $\hat{\bf b}$ is defined to lie in the plane of $\hat{\bf e}$ and $\hat{\bf w}$,
we may define an angle $\eta$ by
 \begin{equation}
 e^{i \eta} = \hat{\bf b} \cdot ( \hat{\bf e} + i \hat{\bf w}) \, .
 \end{equation}
 If we again assume, for simplicity, that $\hat{\bf w}$ is oriented in the $x-y$ plane, at an angle $\alpha$ relative to the $x$-axis, and that $\hat{\bf e}$ is parallel to $\hat{\bf z}$,
we can rewrite the Hamiltonian in Eq.~(\ref{eq:h0}) as:
\begin{eqnarray}
h_0&=& \frac{1}{2m} p^2-\mu + u p e^{- i \sigma_z \alpha/2} \sigma_y e^{i \sigma_z \alpha/2}  \nonumber \\
&+&B e^{- i \sigma_z \alpha/2} e^{- i \sigma_y \eta/2} \sigma_z  e^{ i \sigma_y \eta/2} e^{i \sigma_z \alpha/2} \label{eq:brot}
\end{eqnarray}
This will clearly lead to an additional rotation of the Majorana spinor.  If we generalize these formulas to the case where $\hat{\bf w}$ and $\hat{\bf e}$ are reoriented by an arbitrary SO(3) rotation, we find
 \begin{equation}
\hat{U} =  e^{i \theta / 2} S  e^{-i \sigma_y \eta/2}
\end{equation}
In order to make $\hat{U}$ single-valued, we shall require $-\pi < \eta \leq \pi$ .  This gives us an additional cut, with a jump in sign when $\eta$ crosses the value $\pm \pi$.

In all cases, if the parameters of the Hamiltonian undergo a continuous change which returns to its starting value, so that ${\bf{\Omega}}_T = {\bf{\Omega}}_0$, then the  matrix $\hat{U}_j$ will also return to its starting value. In doing so, it may have undergone a discontinuous sign change one or more times.  The number $N_s$  of these sign changes, mod 2, will be equal to the sum of the number of times that $e^{i \eta}$ and $\Delta$ wind around the origin, plus one if the accumulated rotation of the pair $(\hat{\bf e}, \hat{\bf w})$  is equivalent to a rotation of $2 \pi$.   We note that the net sign change, $-1^{N_s}$, is a topological invariant of the path in parameter-space. It cannot be changed by any continuous deformation of that path so long as the gap remains finite throughout.  This net sign change will be important in our analysis of the physical results of any manipulations of the Hamiltonian parameters.

To conclude, in this section we explored several topological manipulations on the two Majoranas at the end of a topological segment. Some of these manipulations do not involve any motion of the Majoranas themselves. We showed that a rotation of the triad $\hat{\bf e},\hat{\bf w},\hat{\bf e} \times \hat{\bf w}$ by $360^\circ$ degrees around any axis leads to a multiplication of the Majorana operators by -1, provided the direction of the Zeeman field is simultaneously rotated so that it is fixed with respect to triad. As a particular case, if the magnetic field is kept parallel to the wire, and the wire is rotating around its own axis, so that $\hat{\bf e}$ is rotated by $2 \pi$ while $\hat{\bf w}$ and $\hat{\bf b}$ are fixed, then the two Majoranas are multiplied by $-1$. Similarly, a change of the bulk $s$-wave superconducting phase, for example by taking a vortex around the wire, will lead to a multiplication of the Majoranas by -1. Additional factors of $-1$ may occur if the orientation of the Zeeman field is rotated relative to the triad, in such a way that the projection of the Zeeman field on the $\hat{\bf e}-\hat{\bf w}$ plane winds  around the origin. (This projection can never be too small or one would lose the gap in the wire). Majoranas can also be manipulated by depleting sections of a wire and pushing the Majoranas around bends. In section \ref{se:BasicY}, we shall see how Majoranas may also be manipulated by pushing them through Y-junctions.

\section{\label{se:onewire} Adiabatic manipulation of Majorana operators}

\subsection{\label{se:genprin}General principles}

In the previous section we introduced the zero energy Majorana states (or modes) at the two ends of each wire in the two models we consider. In this section we explain the methods with which we analyze manipulations involving these modes. We first consider the general principles that determine the adiabatic time evolution of Majorana fermion operators. Then, we relate the Majorana operators to physical observables, and emphasize their relation to the parity of the number of electrons in each wire. Subsequently, we apply the general methodology to manipulations of $N=1$ and $N=2$ wires.

Note that when analyzing the time evolution of the system under a time dependent $\Omega$ we will not limit ourselves to periodic trajectories. In a typical physical realization the requirement of a precisely
periodic variation of the parameters, i.e., $\Omegas_T=\Omegas_0$, is rather stringent. When a vortex encircles  a wire, or when a wire is
rotated around its center, it is unphysical to require that all atoms involved in the motion return precisely to their original
positions. Rather, we shall make the much less restrictive assumption that the variation of $\Omegas$ will be such that the Majoranas are always very far from each other,  and  the
number of zero energy Majorana modes in the system is never changed. Under this assumption the Majorana operators $\gamma_i$ keep their identity throughout their
motion, and it is meaningful to track the time evolution of expectation values of products of Majorana operators such as $\left\langle \gamma_i \gamma_j\right\rangle$ or $\left\langle\gamma_i \gamma_j \gamma_k \gamma_l \right\rangle$.

According to Ehrenfest's theorem, if $\left|\Psi(t) \right\rangle$ is the many-body wave function at time $t$, and $A$ is an operator with an explicit dependance on the parameters $\bf \Omega$, then

\be
\label{timeev}
\frac {d  \sbra{\Psi}{\hat A}\sket{\Psi}}{dt} = \frac {\left\langle {\Psi }\left| [ \hat A,H ] \right|\Psi\right\rangle}  {i} +
  \left\langle {\Psi} \left|{  \pa {\hat A}{\Omegas}}\right|{\Psi} \right\rangle \cdot \frac {d \Omegas}{dt}
\ee
If $\hat A$ is a product of zero energy Majorana operators corresponding to the parameters $\bf \Omega$, then $[\hat A, H ]=0$, and the first term is absent. Remarkably, the second term is also zero when $\Psi$ is in the ground state manifold, except at isolated points in the trajectory where $\left\langle \Psi \left| \hat A \right| \Psi \right\rangle $ discontinuously changes its sign, because of our sign convention.  Other than these isolated points, the expectation value of $\hat A$ is time independent. This statement does not assume that  the Majoranas involved belong to the same wire segment or to a single  cluster of occupied segments connected by $Y$ junctions,  as long as they are connected to the same bulk superconductor, so that the superconducting phase is well defined.


To see that this is true, we first note that within the BdG model, we may write $\partial \gamma_j / \partial \Omegas = \sum_l  c(\Omegas)_l \gamma_l +  {\bf Q}$, where $\bf Q$ where Q is a linear superposition of finite-energy Bogolyubov quasiparticle operators.  If $\gamma_j$ is located spatially far from all the other Majorana operators, so that their BdG spinors do not overlap, then the coefficients $c(\Omegas)_l$ must vanish for all $l \neq  j$.   Moreover, if the parameters $\Omegas$ are not at a point where the  wave function  for $\gamma_j$ jumps discontinuously, then $\{\gamma_j , \partial \gamma_j/\partial\Omegas \} = 0$, since $\gamma_j^2=1$.   Thus we also have $c(\Omegas)_j=0$. (See Supplementary Material of Ref.\ [\onlinecite{Alicea10}] for more details.)  On the other hand, if $\Omegas$ crosses a special point $\Omegas_1$ where the BdG wave function for $\gamma_j$ has a discontinuity in sign, then $\gamma_j$ will also  undergo a discontinuity in sign, without affecting the relation $\gamma_j^2=1$. (Obviously, the derivative of $\gamma_j$  is not well defined at such a  point.) Combining these results, we see that the expectation value $\left\langle{\Psi} \left|\pa{\hat A}{\Omegas}\right|{\Psi}\right\rangle$ is zero at all points except the points where the operator $\hat A$ changes sign due to a sign change of one of Majorana operators out of which it is composed. Notice that this conclusion is valid beyond the quadratic model of the BdG equation,  as long as the ground state subspace is spanned by hermitian Majorana operators $\gamma_j$ that do not spatially overlap with one another.

This observation leads to the conclusion that the set of parity states to which the system may be brought by topological manipulation is rather limited.  To understand why it is so, assume that at the beginning there is a completely depleted network of wires. Then an application of gate potentials forms topological superconducting sections with Majoranas at the end points. Assuming that during the application of the gate potentials there was no coupling between the sections we conclude that the parity of each section is even.

 Performing now topological manipulations such as braiding of Majoranas we can follow the evolution of the operator $\hat A =\gamma_i \gamma_j$. As argued earlier the only possible operation is a multiplication of the operator observable $\left\langle A  \right\rangle \equiv \left\langle \Psi \right| A \left|\Psi \right\rangle$ by a minus sign at the discontinuous points. For example if $\gamma_1$ and $\gamma_2$ initially share a common section then they are in a pure even state with $i\left\langle \gamma_1 \gamma_2 \right\rangle =1$. A braiding operation of $\gamma_1$ with $\gamma_2$ keeps  the parity $1$. Double braiding of $\gamma_2$ with an additional Majorana $\gamma_3$ will change the parity to $-1$. A single braiding of $\gamma_2$ with $\gamma_3$ leaves the wire section with $\gamma_1$ and $\gamma_3$ at its ends. The initial value $\left\langle \gamma_1 \gamma_3 \right\rangle=0$ acquires a minus sign but remains zero. Hence the wire section must be in a mixture of either even or odd parities with equal weight.
 Generalizing this example for any pair of Majoranas we conclude that after the topological manipulation each wire section  may end up in a pure state of even and odd parity, or in a mixture of even and odd parities, at equal weight.

These observations point our attention to the important, yet confusing, issue of sign discontinuities in the Majorana operators $\gamma$. We shall address this issue in detail, introducing two possible descriptions, one that avoids discontinuities at the price of multiple-valuedness, and one that uses discontinuities to ensure single-valuedness. In this discussion we will also relate the expectation values of Majorana operators to the physically interesting observable - the parity of the number of electrons in each wire. In an attempt for a clear exposition of the two descriptions of this problem, we start with a simpler, but related, problem: a spin $1/2$ in a magnetic field.

\subsection{Spin in a time-dependent magnetic field}


We consider two quantum states of  spin with $s=1/2$ in a magnetic field lying in the $x-y$ plane, whose direction forms a time-dependent angle $\alpha(t)$
with the $x$--axis. We analyze the phase the spin accumulates when $\alpha(t)$ winds adiabatically by $2\pi$. The corresponding spinors have equal probability for spin $\sigma_z = \pm 1$, which is analogous to the equal weight of the electron and the hole in Majorana fermions. The spin is described by the Hamiltonian
\begin{equation}
H_\sigma=-B(\sigma_x\cos{\alpha}+\sigma_y\sin{\alpha})
\label{spin12}
\end{equation}
with $B$ and $\alpha$ being parameters. The instantaneous eigenstates of $H_\sigma$ depend
only on the (fixed) $\alpha$, and the ground state in the $\sigma_z$ basis takes the form
\begin{equation}
\left|{\rm gs}(\alpha)\right\rangle= e^{i\chi(\alpha)}\frac{1}{\sqrt{2}}\left (\begin{array}{c}e^{i\alpha}\\ 1\end{array}\right )
\label{spings}
\end{equation}
where $\chi(\alpha)$ may be chosen arbitrarily. When $\alpha$ varies slowly in time with initial angle $\alpha_0$, a spin that is initialized in the instantaneous ground state $\left|{\rm gs}(\alpha_0)\right\rangle$ evolves in time into $e^{iBt-i\theta_B}|{\rm gs}[\alpha(t)]\rangle$. The accumulated phase includes a dynamical part, $Bt$, a Berry phase part,
\begin{equation}
\label{eq:Berry}
\theta_B[\alpha(t)]={\rm Im}\int_{\alpha_0}^{\alpha(t)} d\alpha\langle {\rm gs}(\alpha)\left|\partial_\alpha {\rm gs}(\alpha)\right\rangle,
\end{equation}
and the contribution $\chi\left[\alpha(t)\right]-\chi\left[\alpha(0)\right]$ emerging from the explicit dependence of $\left|{\rm gs}[\alpha(t)]\right\rangle$ on $\chi[\alpha(t)]$. The difference $\chi_{\rm mon}=\chi(\alpha_0+2\pi)-\chi(\alpha_0)$ is, by definition, the monodromy phase. Notice that while the adiabatic evolution of the ground state, including its accumulated phase, is uniquely determined by the Schr\"odinger equation, the way the geometric phase is split between the Berry phase and the monodromy depends on the choice of
$\chi(\alpha)$.

Indeed, a straightforward calculation of the Berry phase $\theta_B(\alpha_0+2\pi)$, using Eqs.~(\ref{spings}) and (\ref{eq:Berry}), yields
\begin{equation}
\label{thetaB}
\theta_B(\alpha_0+2\pi)= \pi+\chi(\alpha_0+2\pi)-\chi(\alpha_0).
\end{equation}
When computing the full geometric phase, $\chi_{\rm mon}-\theta_B(\alpha_0+2\pi)$, the last two terms in (\ref{thetaB})  are canceled by the monodromy, resulting in a geometric phase of $\pi$, independent of the choice of $\chi(\alpha)$.

Similar to gauge fixing, the choice of $\chi(\alpha)$ is a matter of convenience, simplicity, and clarity of the calculation or argumentation. Three choices are worth discussing in the present context. If one chooses $\chi(\alpha)$ to satisfy $\chi(\alpha+2\pi)=\chi(\alpha)$, the instantaneous ground state is single-valued as a function of $\alpha$ and the total accumulated phase is given by the Berry phase $\theta_B[\alpha_0+2\pi]$. In contrast, choosing $\chi(\alpha)=-\alpha/2$ makes the two components of the instantaneous ground-state spinors (\ref{spings}) complex conjugates of one another, not unlike spinor representations of Majorana fermion operators. For this multi-valued representation, the Berry phase vanishes, $\theta_B[\alpha(t)]=0$, so that the geometric phase is accumulated entirely through the monodromy.

\begin{figure}[h] \begin{center}
 \includegraphics*[width=.9\columnwidth]{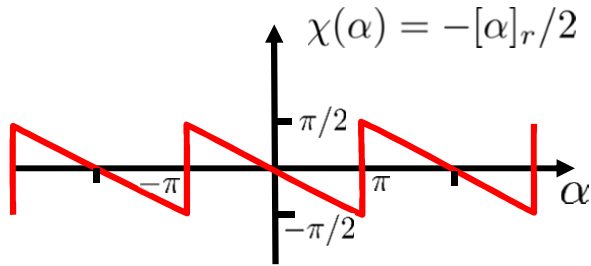}
  \end{center} \vspace{-.5cm} \caption{\label{fg:chi} Plot of the function $ - [\alpha]_r / 2$, where $[\alpha]_r \equiv [(\alpha+\pi)\!\! \mod 2\pi]-\pi$; cf. Eqs.~(\ref{eq:modx}) and (\ref{eq:modx1}). Choosing $\chi(\alpha)=-  [\alpha]_r / 2$  in Eq.~(\ref{spings}) we have a single-valued Majorana-like representation of the ground state.} \end{figure}

Finally, it is also possible to construct a Majorana-like representation which is {\em single-valued} by choosing $\chi(\alpha)= - [\alpha]_r / 2$, where $[\alpha]_r $ is defined according to Equations (\ref{eq:modx}) and (\ref{eq:modx1}); cf. Fig.~\ref{fg:chi}.
For this choice, $\chi(\alpha)=\chi(\alpha+2\pi)$, so that the total geometric phase originates from the Berry phase $\theta_B(\alpha_0+2\pi)$ only. However, the Berry phase is  now accumulated in discontinuous jumps of $\chi(\alpha)$ at $\alpha=\pi (2n+1)$ with $n$ being an integer. We note that the time evolution of the physical wave function remains continuous, even for this discontinuous choice of $\chi(\alpha)$

\subsection{Application to Majorana variables}
\label{se:AppMaj}

As in this spin-1/2 example, there is some freedom in defining the instantaneous eigenspinors $\psi$ that correspond to the Majorana fermion operators, and there are analogs to both the single-valued and multi-valued choices we defined above.

A zero-energy solution $\psi$ of the BdG equation that starts in Majorana form $\left(\psi_p,{\cal T}\psi_p\right )^T$ at $t=0$ stays in that form throughout its time evolution, and satisfies $\langle\psi|\partial_t\psi\rangle=0$ at all times. As a consequence, upon completion of a periodic trajectory, $\Omegas_0=\Omegas_T$, both the zero-energy Majorana spinor $\psi$ and the corresponding Majorana operator $\gamma$ must return to their initial values {\em up to a possible sign}. As before, if the Majorana operators are defined singly-valued as a function of $\Omegas$, they acquire discontinuous minus signs at particular values of $\Omegas$. If continuity is preferred and the price of multi-valuedness is paid, then for every value of $\Omegas$ the Majorana operators are defined only up to a minus sign, and these minus signs depend on the entire history $\Omegas(t)$.

\subsubsection{Electron number parity.}
Let us now relate this choice of Majorana operators to physically measurable quantities, in particular to the electron parity. We start with the simplest case, that of a single wire, with Majorana modes $\gamma_1,\gamma_2$ at its two ends. For such a wire we may define an electron parity operator $P$, which may take the values $\pm 1$, depending on whether the number of electrons in the wire segment is even or odd.
We naturally expect that  within the zero-energy Hilbert space, the parity operator should be related to  $i\gamma_1\gamma_2$, with the eigenvalues being $\pm 1$. Far less clear, and far more confusing, is the assignment of the two eigenvalues to even and odd number of electrons, and the way this assignment evolves when $\Omegas$ varies in time.

Generally, we may write
\begin{equation}
P = i k \gamma_1 \gamma_2 \, ,
\end{equation}
where $k=\pm 1$.  The value of $k$ may depend on the orientations of the wire ends and the sign conventions used in the definition of the Majorana operators.
It is easy to see that different choices for the dependence of $\gamma_1,\gamma_2$ on $\Omegas$ lead to different values for $k$. We note, however, that the actual electron number parity cannot change in manipulations of a single wire, as the parity is an invariant quantity in any adiabatic manipulation  that does not bring it into contact with another wire and does not deplete the wire to the point where all electrons are expelled.  In turn, this reflects the fact that the tunnel coupling to the bulk superconductor allows only electron pairs to tunnel between the superconductor and the wire,  unless one supplies enough energy to inject an electron or hole above the energy gap of the bulk superconductor.

 Consider a straight  wire lying in the $x-y$ plane, at an angle $\phi$ relative to the $x$-axis, with $\hat{\bf e}$ and $\hat{\bf b}$ in the z-direction. We define angles $\alpha_j$  as the azimuthal angle relative to the $x$-axis of a  line parallel to the wire at  end $j = 1,2$, directed from the depleted region into the  region occupied by the topological superconductor, with the restriction $-\pi \leq \alpha_j < \pi$.  Then, $\alpha_1=\phi$ and $\alpha_2=\phi\pm \pi$. The Majorana operators $\gamma_1,\gamma_2$ depend on $\alpha_1,\alpha_2$ respectively, and that dependence may be chosen to be either single-valued or multi-valued with respect to $\alpha_j$.
We imagine now that the wire is rotated by an angle $\pi$  around its center, and examine what happens to $P,k,\gamma_1$ and $\gamma_2$.
This is a point where the difference between the single- and multi-valued representations should be carefully followed.

If the Majorana operators are defined in a way that is single-valued with respect to $\alpha_j$, then the interchange of $\alpha_1$ with $\alpha_2$ leads to the interchange $\gamma_1\rightarrow\gamma_2$ and $\gamma_2\rightarrow\gamma_1$. That is, after the interchange, the expression for $\gamma_1$ in terms of the electron creation and annihilation operators is the same as the expression for $\gamma_2$ before the interchange and vice versa. Since  $i \gamma_2 \gamma_1 = - i \gamma_1 \gamma_2$ and since the number parity is not changed by the rotation, the pre-factor $k$ has necessarily changed sign. That change of sign takes place at the value of $\phi$ where one of the $\gamma$'s acquires a discontinuous sign change. The parity operator as expressed in terms of creation and annihilation operators for the electrons  does not change  discontinuously at this point. Similarly, the actual physical value of the parity does not change during the manipulation.

In contrast, if the Majorana operators are defined in a way that is multi-valued with respect to $\alpha_j$, they generally depend on $e^{\pm i\alpha_j/2}$. Then, a rotation by $\pi$ leads to the transformation $\gamma_1\rightarrow\pm\gamma_2$ and $\gamma_2\rightarrow\mp\gamma_1$, with the proper sign being determined by the sense of the rotation. With that choice, the parity operator stays $P=i\gamma_1\gamma_2$ and $k$ does not change. Again, no discontinuous change occurs in the parity operator when expressed in terms of electrons creation and annihilation operators or in the physical value of the parity.

Comparing these two descriptions, the advantage of the single-valued choice of the $\gamma$'s is that all quantities are fully determined by the instantaneous values of the parameters, in this case $\alpha_j$, and are independent of the values of the parameters in earlier times. The dependence on the history of the system - preparation, manipulation etc. - is all in the many-body wave function of the wire. These advantages come at the price of having discontinuities in the $\gamma$'s and in the way $P$ is expressed in terms of the $\gamma$'s. These discontinuities appear in arbitrarily chosen locations in parameter space. In contrast, the advantage of the multi-valued way is in the absence of any of these discontinuities, while the disadvantage is in an arbitrariness in the initial definitions of the operators $\gamma_1,\gamma_2$ and the need to follow the time evolution of these operators to be able to relate them to the parity $P$.

The two descriptions may be extended to more complicated situations. Generally, any $N=1$ wire may be obtained  from an $N=1$ straight wire by continuous bending. In the single-valued description, as the wire is bent from being straight to its final form, sign discontinuities may occur in the Majorana operators $\gamma_1,\gamma_2$. These discontinuities occur when either of the angles $\alpha_j$ crosses a cut. Whenever such a sign change occurs, a similar sign change occurs in the prefactor $k$, thus guaranteeing that the parity of the number of electrons remains continuous. It should be noted that the form of the parity operator is not fully determined by $\alpha_j$. Two wires with the same values of $\alpha_j$ may have opposite values of $k$, if the deformation of one into the other involves an odd number of crossings of cuts.

In the multi-valued description,  the prefactor $k$ does not vary at all with the bending of the wire, but the values of the angles $\alpha_j$, of the factors $e^{i \alpha_j/2}$ and hence of the operators  $\gamma_1,\gamma_2$ depend on the trajectory that brings the wire from its initial to its final form.

For the $N=2$ case, two wires may be deformed into one another as well, but the deformation is more complicated, since the Majorana modes depend not only on the directions of the wire's ends, $\alpha_j$, but also on the directions of $\hat{\bf e}$ and $\hat{\bf b}$.
Again, in the single-valued description a discontinuous sign change associated with cuts in the definition of $\gamma_j$ as a function of $\Omegas$ leads to a sign change in $k$, leaving the parity continuous. And again, the value of $k$ is not fully determined by the values of $\alpha_j, \hat{\bf e}$ and $\hat{\bf  b}$ at the two ends. Rather, a variation of these parameters along the wire may lead to sign changes in $k$ relative to its value for a reference straight wire where, say, $\alpha_j=0,\pi$ and ${\hat{\bf e}},{\hat{\bf b}}$ are parallel to the $\hat{\bf z}$-axis.

\subsection{Examples}\label{examples subsection}

We now analyze two $N=1$ examples, of a straight and a curved wire. For both, we calculate the parity operator in a representation that is single-valued with respect to $\Omegas$. Following that we will discuss in subsection \ref{subse:N2} the $N=2$ case.
\subsubsection{Straight wire}
To find the parity operator for a straight wire we use the discrete version of the Hamiltonian of Eq.~(\ref{pwave}) with $t=\mu$, $a^2=1/(m \mu)$, $\Delta=v/a$ and the number of sites $N=L/a+1$. Here with $a$ being the lattice constant and $L$ the wire length, the Hamiltonian becomes
\begin{equation}
H=-t\sum_{l=1}^{N-1} c^\dagger_l c_{l+1}
-|\Delta| e^{i (\theta + \tilde{\alpha})} c^\dagger_l c^\dagger_{l+1}+h.c ,
\label{H1d}
\end{equation}
where $\tilde{\alpha}$ describes the orientation of the wire in the $x-y$ plane, directed from the end $l=1$ to the end $l=N$.  Since, in our conventions for Majorana  operators at the end of a wire, we have defined the angle $\alpha$ to describe the orientation of a vector pointing into the wire from the end, we note that $\alpha = \tilde{\alpha}$ for a Majorana at the end $l=1$, but $\alpha = \tilde{\alpha} \pm \pi$ at the end $l=N$.

Let us now define
\begin{eqnarray}
a_l&=& e^{i \chi/2} c_l+ e^{-i \chi/2} c^\dagger_l, \nonumber \\
b_l &=&i(e^{i \chi/2 } c_l - e^{-i\chi/2}   c^\dagger_l), \; \;
\chi = - [\theta + \tilde{\alpha}]_r \, ,
 \label{eq:beta}
\end{eqnarray}
which gives
\begin{equation}
\begin{array}{cclcc}
i a_l b_l &=& 1-2 c_l^\dagger c_l &&\\
\left\{c^\dagger_l, c_l\right\}&=&1 &\Rightarrow& a_l^2=b_l^2=1\\
\left\{a_l, a_m\right\}&=&\left\{b_l, b_m\right\} = 2\delta_{lm}, && \left\{b_l, a_m\right\}=0.
\end{array}
\end{equation}
If we now consider the case
  $|\Delta| = t > 0$,  the Hamiltonian may be rewritten as:
\begin{equation}
\label{eq:kit}
H=t\sum_{l=1}^{N-1}  i a_{l+1} b_l
\end{equation}
The Hamiltonian clearly commutes with $a_1$ and with $b_n$. These two operators are therefore the  Majorana end modes, up to possible factors of $\pm 1$. In order to determine the sign, we compare the definitions in (\ref{eq:beta}) with the sign conventions defined for the continuum wave function  (\ref{eq:majneq1}).  We see that the Majorana operator at $l=1$ may be written as $\gamma_a = a_1$, but  at the other end, we have
\begin{equation}
\gamma_b = b_N \,  {\rm {sgn}} \left([\theta + \tilde{\alpha } ]_r  \right)\, .
\end{equation}

In terms of the Majoranas,  the parity operator is
\begin{equation}
P=\prod_{l=1}^N \left(1-2 c_l^\dagger c_l\right) = \prod_{l=1}^N \left(i a_l b_l\right).
\end{equation}
When the wire is empty all sites are empty and the parity is 1.
In the ground states of Hamiltonian~(\ref{eq:kit}) $\left\langle i b_l a_{l+1}\right\rangle =1$ so that for the ground state manifold the parity operator is
\begin{equation}
\label{eq:parity}
P=ia_1 b_N \prod_{l=1}^{N-1} \left(i b_l a_{l+1}\right) = i a_1 b_N =  i k \gamma_a\gamma_b ,
\end{equation}
\begin{equation}
\label{kstraight}
k =  {\rm {sgn}} \left([\theta + \tilde{\alpha } ]_r  \right)\, .
\end{equation}

Although we have explicitly considered only the special case where $t = |\Delta|$ and the chemical potential is at zero,  the  relation between $P$ and $i \gamma_a  \gamma_b$ cannot change discontinuously if the parameters are varied, as long as there is an energy gap in the bulk of the wire, and the orientation of the wire is fixed. Furthermore, when the chemical potential is not fixed at the center of the band, we can take the continuum limit of $a \to 0$, with the electron density held fixed.  This recovers the continuum $N=1$ model defined earlier.

\subsubsection{Curved wire}
To determine $k$ explicitly for a curved wire in two dimensions we have to follow the configuration of the wire in the two dimensional space. Assuming that  the curve describing the wire is $(x(s),y(s))$ we define an angle $\tilde{\alpha}(s)$ as the orientation of the tangent vector at  point $s$, choosing  the vector to point along the wire direction leading from a specified end ($s=0$) to the other, and requiring that  $\tilde{\alpha}$ is a continuous and slowly varying (on the coherence length scale) function of $s$. We see that
$\tan \tilde {\alpha} = dy/dx$ and the curvature $\kappa =  d\tilde{\alpha} /dw = (x'y''-y' x'')/(x'^2+y'^2)$ with $w$ being the arc length coordinate along the wire, defined by $dw=\sqrt{x'^2+y'^2} ds$. We define the net orientation change as  $\Delta \tal = \tal(L) - \tal(0) = \int_0^L \kappa \, dw$, where $L$ is the length of the wire segment and  we have chosen the parameter $s$ to be the arc length $w$.
Note that $| \Delta \tal |$  can be larger than $2 \pi$ if there are loops in the wire.   We shall assume here that the phase $\theta$ of the bulk superconductor is a constant along the wire.

If $| \Delta \tal |$ lies in the interval $2\pi (M-\frac{1}{2}) < | \Delta \tal | < 2 \pi (M + \frac{1}{2})$, (with M being an integer), then our result for the factor $k$  relating the electron number parity operator to the product $i \gamma_a \gamma_b$ is
\be
\label{kcurved}
k =   (-1)^M  q \, , \, \, \, \, q \equiv   {\rm {sgn}} \left([\theta + \tilde{\alpha }(L) ]_r - [\theta + \tilde{\alpha }(0)- \pi ]_r
  \right)\, .
\ee
To derive this result, consider what would happen if one were to adiabatically change the parameters in the Hamiltonian so that the wire segment becomes straight, while holding fixed the orientation $\tal(0)$ at the first end.  The wire orientation $\tal(L)$ at the second end will change continuously from its initial value $\tal_0(L)$ to a final value equal to $ \tal(0)$.  Of course the expectation value of the electron parity will  be unchanged in the procedure.

During the process of straightening, the value of $M$ will change by $\pm 1$  whenever $\tilde\alpha(L)-\tilde\alpha(0) = (2 n + 1) \pi$, but the sign of $q$ will also change at these points.  The sign of  $q$ will also change when $\phi+\tilde \alpha(L) = (2 n+1) \pi$, but the operator $\gamma_b$
at the end $w=L$ will also change sign at such points. Therefore the product $i k \sex {\gamma_a \gamma_b}$ will be unchanged during the process.  When the wire is straight, however, we have $\Delta \tal = 0$, and we see that $[\theta + \tilde{\alpha }(L) ]_r $ and $[\theta + \tilde{\alpha }(0) - \pi]_r$ have opposite signs.  Therefore  (\ref{kcurved}) reduces to (\ref{kstraight}).

\subsubsection{Wires with  $N=2$}
\label{subse:N2}
For $N=2$ the Majoranas depend on a richer set of parameters, and thus cross more types of branch cuts. For example, cuts may occur if we consider situations where the direction of $\hat{\bf e}$ is allowed to change along the wire, or if the wire orientation is lifted out of the plane, or if the angles $\eta$ or $\theta$ are allowed to vary along the wire.   In these cases we must follow all the orientation parameters  along the wire.

To describe the general situation, we begin by defining the unit vectors  $\hat{\bf e}, \hat{\bf b}, \hat{\bf w}$, at all points along the wire, where $\hat{\bf w}$ points in the wire direction from end $a$ to end $b$.  We require that $\hat{\bf e}$ is perpendicular to $\hat{\bf w}$ at  all points, and we define $\eta(w)$,  as before, as the angle between $\hat{\bf e}$ and the projection of $\hat{\bf b}$ onto the plane formed by $\hat{\bf e}, \hat{\bf w}$.  We define an SO(3) matrix $R(w)$ which describes the orientation of
 $\hat{\bf e}, \hat{\bf w}$ relative to the laboratory axes  $ \hat{\bf x},\hat{\bf y},\hat{\bf z}$.
 We may define angles $\Delta \eta$ and $\Delta \theta$ to describe the net change in $\eta$ and $\theta$ along the wire, when these angles are defined to vary continuously as a function of $w$.  We may also define an SU(2) matrix $S(w)$ as the matrix corresponding to the relative rotation  $R(w)R^{-1}(0)$, with the requirement that $S(w)$ is a continuous function of $w$.   With the sign conventions for $\gamma_a$ and $\gamma_b$ specified earlier,  we  find that the expression for $k$ has the form
 \be
 k = (-1)^M Q  \,\, {\rm sgn} \left[{\rm tr} S(L) \right] \, ,
 \ee
\be
M = {\rm Int} (\Delta \eta /  (2 \pi)) +  {\rm Int} (\Delta \theta /  (2 \pi)) \, ,
\ee
where Int$(x)$ denotes the closest integer to $x$,  and $Q= \pm 1$ depends on the matrices $R(0)$ and $R(L)$.  Although we have not found a simple form for the function $Q$, the essential point is that its value may be computed if desired in any instance, and it  depends only on the end points of the orientations and not on the path between them.

To derive the result we may again begin  by considering a straight wire in the $x-y$ plane, with $\hat{\bf b}$ independent of position and  $\hat{\bf e} = \hat{\bf z}$.  We again define the angle $\tal$ as the orientation of  the wire directed from the end with Majorana operator $\gamma_a$ towards the end with $\gamma_b  $.  If we now assume that we are in the strong-field limit,  we can map the problem onto the $N=1$ Kitaev model considered previously.  The constant $k$ in this case is given by
\be
\label{kstraight2}
k =       {\rm {sgn}} \left([ \tilde{\alpha } ]_r  \right)\, .
\ee
The difference between this expression and (\ref{kstraight}) arises from the different convention we have adopted for the branch cuts in the two cases.  In the present $N=2$ case,  the jump in sign of a Majorana variable occurs when the orientation of $\alpha$ (pointing into the wire from the end) crosses $\pi$ mod $2\pi$, independent of the phase of the superconducting order parameter.  The result is also independent of the angle $\eta$.

The results for a bent wire of arbitrary orientation, and with $\eta$ and $\theta$ that vary along the wire, may be obtained by deforming the wire adiabatically into the case  of a  straight wire in the $x-y$ plane.  The value of  $k$ will be multiplied by $-1$ each time the orientation parameters of the Majorana at either end of the wire cross a discontinuity surface.

As a particular example, consider a bent wire in the $x-y$ plane, with $\hat{\bf e} = \hat{\bf z}$ everywhere.  Suppose that $\tal (w)$ describes the local orientation of the wire, as in the subsection on $N=1$,  and suppose the total bending angle $\Delta \tal$  is less than $2 \pi$.  In this case we find
\be
\label{kcurved2}
k = {\rm sgn} \left( [\tal (0)]_r - [ \tal(L) - \pi]_r \right) = {\rm sgn} \left( [ \alpha_a  ]_r - [\alpha_b]_r \right) \, ,
\ee
where $\alpha_a$ and $\alpha_b$ describe the orientations of the Majoranas at the two ends.  This result coincides with the result (\ref{kcurved}) for the $N=1$ case when the phase $\theta$ of the superconducting order parameter is zero.

\section{Moving Majoranas through a  $Y$ Junction}
\label{se:BasicY}
In the previous section we analyzed two Majorana operators at the ends
of a single wire, and showed how their states may be manipulated in
such a way that preserves their identities but implements non-trivial
transformations. In this section we enlarge the system to include
several wire segments connected by junctions, which will be  necessary in order to perform non-trivial
transformations on the state of the system, and in particular, braiding.

Initially, our system consists of an even
number of Majoranas, distributed into pairs, with each pair consisting
of end modes of the same topological segment. Any non-trivial
transformation will require a rearrangement of the pairing configurations while keeping the
identities of the Majorana states distinct, so that the
quantum information encoded in their state is preserved. Thus, the prototypical situation we will eventually consider is one in which at $t=0$ we have two topological segments, with the Majorana operators $\gamma_a,\gamma_b$ located at the two ends of one segment and $\gamma_c,\gamma_d$ located at the ends of the second segment.  At the final time  $t=T$,  one segment has $\gamma_a,\gamma_c$ at its ends and the other has $\gamma_b,\gamma_d$ at its ends. This exchange of the two Majoranas should be carried out through an adiabatic evolution of the parameters ${\bf{\Omega}}$, without introducing a coupling that will modify the energy of any of the Majorana states.
The analysis of such manipulations requires care, as exemplified by the following scenario. If both wires are initially in a state of even parity, the interchange of $\gamma_b$ with $\gamma_c$ will result in a state where the parity of both wires has a zero expectation value. If the interchange is then repeated, the two wires may end up both at even parity, or both at odd parity, depending on the precise details of the interchange process. It is the analysis of this process which we now outline.    We shall accomplish this process using a series of elementary manipulations. The elementary tool for these
manipulations is the $Y$ junction. At the  $Y$ junction three wires come together each of which may be ``empty"  or ``full" ({\it i.e.}, in a TS state).  If the number of TS segments at the junction is odd, there will generically be one zero-energy Majorana state at the junction.  If the number of wires is even, there will be no such state.

The case of three full (i.e., topological) wires may be understood by depleting each wire just before it reaches the junction.
If tunneling between the wire ends were completely forbidden, there then would be a zero-energy Majorana state at the end of each wire, {\it i.e.}, three Majoranas all together.  However, when tunneling is allowed, the three states will be split, with two combinations of the original Majorana operators pushed to finite energy, and one combination forming a new Majorana operator with precisely zero energy.

To see this formally, we use the Hamiltonian describing the tunneling between Majoranas $1$,$2$, and $3$ at the ends of the three wires near the $Y$ junction, $H=\frac{i}{2}\sum_{l,m=1,2,3} t_{lm} \gamma_l \gamma_m $. The real matrix elements $t_{nm}$ present the coupling between the Majoranas. Since $H=H^\dagger$ we require that $t_{lm}=-t_{ml}$. Writing the solution for the equation $\left[H,\Gamma_n\right]=-E_n \Gamma_n$ as $\Gamma_n=
v_{n1} \gamma_1+v_{n2} \gamma_2+v_{n3} \gamma_3$, we obtain the matrix equation for the amplitudes  $v_{nm}$:
\begin{eqnarray}
&&
i\left(
  \begin{array}{ccc}
    0 & t_{12} & t_{13} \\
    -t_{12} & 0 & t_{23} \\
    -t_{13} & -t_{23} & 0 \\
  \end{array}
\right)
\left(
         \begin{array}{c}
           v_{n1} \\
           v_{n2} \\
           v_{n3} \\
         \end{array}
       \right)=
       E_n \left(
         \begin{array}{c}
           v_{n1} \\
           v_{n2} \\
           v_{n3} \\
         \end{array} \right).
 \end{eqnarray}
 This equation has two particle-hole symmetric solutions with energies $\pm  \sqrt{t_{12}^2+t_{23}^2+t_{13}^2}$, and one zero energy solution corresponding to the true Majorana located near the junction. The zero energy solution is $v_{0m}\propto \epsilon_{l k m} t_{lk}$ (with $\epsilon_{lkm}$ the Levi-Civita symbol). Explicitly, its normalized form is
 \begin{eqnarray}
        && \vec v_0 =\frac{1}{
        \sqrt{t_{12}^2+t_{23}^2+t_{13}^2}}
\left(
 \begin{array}{c}
           t_{23} \\
           -t_{13} \\
           t_{12} \\
         \end{array}
         \right).
\end{eqnarray}

 If the tunnel coupling between two of the wire ends is much stronger than the coupling to the third (e.g. $t_{12}
\gg t_{23},t_{13}$), then, as can be seen explicitly by examining the vector $v_0$ in this limit, the zero energy Majorana operator will be primarily located at this third wire end, while the first two wires combine to form, effectively, a single wire, with a finite energy fermion state inside the superconducting gap of the TS sections.

 \begin{figure}
 \begin{center}
\includegraphics*[width=\columnwidth]{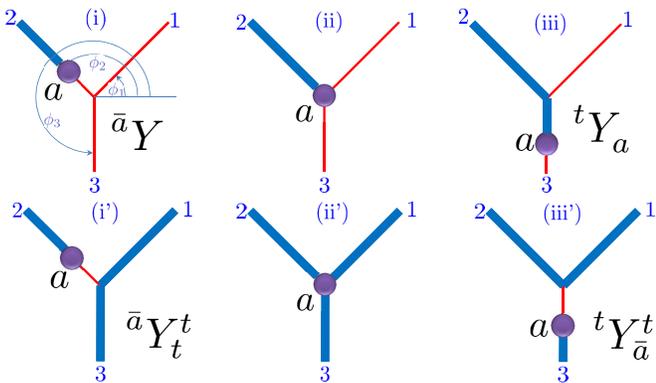}
 \end{center} \vspace{-.5cm}
\caption{ { \label{fg:BasicY}
(Color online) Moving a Majorana across a Y junction.
(i) Basic set up of a $Y$ junction. The bold (blue) line  represents a topological superconductor (TS)  sector while
the thin (red) line denotes an empty wire. The three branches of the junction, $1$, $2$ and $3$ are at angles $0\le\phi_1, \phi_2, \phi_3<2\pi$ respectively to the $x$ axis. (For the N=1 case these are also the phases of the effective $p$-wave superconductor). In this panel, a  Majorana $a$ is  located on branch $2$. The configuration of the wires near the junction is denoted by the symbol $^{\bar a}\!{Y}$, where the bar symbol above the Latin letter $a$ denotes that it is the end of a TS sector stretching away from the junction.  Majorana $\bar a$ is characterized by the angle $\alpha_{\bar a}=\phi_2$.
 (ii) A basic operation is to move Majorana $a$ from branch $2$ to branch $3$. To do so we first move the Majorana to the $Y$ junction, which is done by adjusting gate voltages to fill the remainder of wire 2 with electrons in the TS state.   %
 (iii)   Next we bring a portion of  wire 3 near the junction into the TS state.  This puts the Majorana on leg 3.
 The symbol $t$  in the $^{t}\!{Y_{a}}$ denotes that the entire branch 2 is topological and has no Majoranas. The bar has been removed from the letter $a$ because the Majorana is now at the end of a TS segment connected to the $Y$ junction. It  is characterized by an orientation $\alpha_a = \phi_3 + \pi$.
Lower panels  (i')-(iii'). Shifting a Majorana from branch $1$ to branch $3$ when both branch $2$ and $3$ are in the TS state.
We  denote this by $^{\bar a}\!{Y_{t}^{t}} {\stackrel{s_{\bar 2 \bar 3}}{\longrightarrow}}  {^{t}\!Y_{\bar a}^{t}}$. As shown in the text, this operation gives an additional sign change relative to the case in the upper panels.
}}
 \end{figure}

The basic manipulations at a $Y$ junction involve pushing a  Majorana state from one wire to another through the junction.  Here we may distinguish two situations, illustrated by the upper and lower panels of Figure \ref{fg:BasicY}.  In the upper panels  we have one full wire coming in from far away (wire 2), while the other two are empty.  Initially, the full wire stops short of the $Y$ junction, and the Majorana state  at the end of the filled region, is located on wire 2.  At the end of the manipulation, wire 2 is completely full, and wire 3 is filled to some finite distance from the junction.  The Majorana, at the end of the filled section,  is now found inside wire  3. This manipulation, which does not involve a coupling between two separate topological segments, is rather simple.

The lower panels describe a more complicated manipulation. In this manipulation we  again transfer a Majorana from wire 2 to wire 3, but here we have the opposite situation, where all three wires are full far from the junction.   In the initial state, we assume there is a depleted region in wire 2 between a point $\vec{r}_2$ and the junction, while wires 1 and 3 are everywhere full.   The Majorana is located at the point $\vec{r}_2$. At the end of the manipulation, the depleted region has been moved to wire 3, and the Majorana is located in wire 3. The manipulation involves a transfer of a Majorana between wires.

As long as we change the parameters of our Hamiltonian at  a rate that is small compared to the energy of the lowest positive energy states that occur in the course of the manipulation, positive energy states will never be occupied, and we may ignore them in our considerations.   For the analysis of the time evolution of the single zero energy Majorana operator throughout the elementary manipulations, we find it convenient to revert to a representation in which the Majorana operator evolves continuously with the parameters of the Hamiltonian. We denote this representation of the Majorana operator by ${\tilde\gamma}(\Omega)$. As discussed in Sec.~\ref{se:genprin}, in this representation, the expectation value of the product of ${\tilde\gamma}(\Omegas)$ with any number of Majorana operators far from the junction will be independent of time.
 We can always choose the sign of $\tilde \gamma(\Omegas)$ to be consistent with our sign conventions  at the beginning of the manipulation, ($t=0)$, when the Majorana is on wire 2.  However, we are not guaranteed anymore that ${\tilde\gamma} (\Omegas) $ is still consistent with the convention at the end of the manipulation ($t=T)$ when the Majorana is  located on wire 3. If the conventions are not satisfied, we will encounter a factor of -1 when we convert ${\tilde\gamma}(\Omegas)$ to a Majorana operator obeying our single valued convention, and these factors of -1 must be included when we compute expectation values of physical operators.

Fortunately, the rules for when one needs a factor of -1 are independent of most details of the $Y$ junction. In the remainder of this section, we show how the rules can be deduced from the requirement that the total electron number parity be conserved,  together with our previously developed relations between the electron parity operator of an occupied  segment and the product of Majorana operators at its ends.

In the following analysis, we discuss the behavior of a network of semiconductor wires with $N=2$. We assume that near the junction the phase of the superconductor and the angle $\eta$ are constant, and the wires are in the $x-y$ plane, with $\hat{\bf e}$  in the $z$-direction.  (Results for any other orientation of the $Y$ junction can be obtained by performing a rotation of the system and counting the number of times the parameter set $\Omegas$ at the location of a Majorana crosses  a discontinuity surface during the rotation.)

\subsection{$Y$ Junction Notation}

We first define a notation to keep track of the various arrangements of TS wire segments and  Majoranas  in the vicinity of a junction, such as those exhibited in the various panels of Fig. \ref{fg:BasicY} .   We mark Majorana states on the branches of the Y with a lower case Latin letter located about a Y symbol, at the appropriate location. If the Majorana state is an 'in' state (it is at the end of a topological segment which stretches away from the intersection), we mark the state with an extra bar.
The presence or absence of a bar leads to a difference of $\pi$ in the value of  the angle $\alpha$ describing the Majorana orientation. When a wire branch is filled in the TS from the junction all the way to ``infinity'', so there is no Majorana near to the junction on that branch, we indicate this by including the letter $t$ on the corresponding branch of the Y symbol.

For example, consider:
\be
\leftexp{\bar a}{Y}.
\ee
which describes the state in Fig. \ref{fg:BasicY}(i). Wires 1 and 3 are empty, while
branch 2 contains the Majorana state $\gamma_a$,  terminating a topological segment that stretches away from the intersection.
 Another example is:
\be
\leftexp{t}{Y_{c}}.
\ee
which encodes a Majorana state on branch 3, which terminates a topological phase that stretches to infinity on branch 2, as the $t$ indicates. This is the situation in Fig.~\ref{fg:BasicY}(iii).

\subsection{Derivation from Number Parity Conservation}

Let us first consider the situation illustrated in the upper panels of   Fig   \ref{fg:BasicY}.  We suppose wire 2 is straight and long but finite, so that it has  an additional Majorana at its far end (not shown in the figure) which we label $\gamma_2$.  The orientation of the Majorana $\gamma_2$ is $\alpha_2 = \phi_2 \pm \pi$, while the original orientation of $\gamma_a$ is  $ \alpha_2^0 =\phi_2$.

In our notation, the Majorana operator keeps the same label $a$ throughout the manipulation.  However, the relation of the operator  to the fundamental electron creation and annihilation operators will be different before and after the manipulation, as the parameters $\Omegas$ will be different.  To emphasize this difference, we shall use the symbol $\gamma'_a$, in this section, to denote the  Majorana operator in the final state in the single-valued convention, and use $\alpha'_a$ to denote the  corresponding orientation. In the final state illustrated in  panel (iii) of Fig   \ref{fg:BasicY}, the orientation of Majorana $a$ will be
$\alpha'_a =  \phi_3 \pm \pi$.

 In the initial and final states,  the number parity operators  are respectively  given, according to (\ref{kcurved2}), by
\be
P_{2, a}^0 =  i k_{22} \gamma_2 \gamma_a \, ,  \,\,\,\,
P_{2, a}^T =  i k_{23} \gamma_2 \gamma'_a \, ,
\ee
\be
\label{kij}
 k_{ij} =
{\rm sgn} \left( [ \phi_i-\pi]_r  -  [\phi_j]_r    \right) \, .
\ee
We now relate the operator $\gamma_a$ to the continuously evolving operator ${\tilde\gamma}(\Omegas)$. Since the expectation value $\left\langle \gamma_2 {\tilde\gamma}(\Omegas) \right\rangle_T$ in the final state is equal to the expectation value
 $\left\langle \gamma_2 \gamma_a \right\rangle_0$ in the initial state, and since the expectation value of the number parity must be the same in both states, $k_{23}\left\langle \gamma_2 {\gamma_a} \right\rangle_T=k_{22} \left\langle \gamma_2 {\tilde \gamma}(\Omegas) \right\rangle_T$.  We see that for $\Omegas$ in the final state,  the operator $\gamma'_a $ is equal to  $s_{\bar{2} 3} {\tilde\gamma}(\Omegas)$, with  $s_{\bar{2}3} = k_{22} k_{23}$.
The product $k_{22} k_{23}$ will  equal -1 if  and only if the  $x$-axis lies within the area between two lines of orientation  $\phi_3$  and $\phi_2 - \pi$, with the restriction that the opening angle have magnitude $< \pi$.

 In accord with the discussion in Section \ref{se:genprin}, we know that if $A$ is any operator constructed as a product of Majorana operators far from the $Y$ junction in question,  the expectation value
 $\sex {A {\tilde\gamma}_a(\Omegas)}$ will not depend on time. We therefore see that
 \be
 \sex {A \gamma'_a} =  s_{ \bar{2} 3}    \sex { A \gamma_a}
 \ee
where the expectation value of the left hand side of the equation is taken at time $T$, and that of the right hand side is taken at time zero.

The inverse of the  above manipulation begins with a TS that fills the entire wire 2 and extends, through the junction,  part way into wire 3, so that there is a Majorana located in wire 3. If we now adiabatically deplete the occupied region of wire 3, along with a region close to the junction in wire 2, we will have moved the Majorana from wire 3 to wire 2.   Clearly the sign change associated with the inverse process, which we denote $s_{3 \bar{2}}$, will be the same as $s_{ \bar{2} 3} $.

When all three wires are in the TS state far from the junction, as in the lower panels of Fig.  \ref{fg:BasicY},  the analysis is slightly more complicated.
 We assume that all wires are straight, and we include two more Majorana operators, $\gamma_1$ and $\gamma_3$ at the end of their respective wires.  In this case, the number parity for individual wires is not conserved, but the total parity is.  The parity operators in the initial and final states are given by
\be
P^0 = - k_{22} k_{31} \gamma_2 \gamma_a \gamma_3 \gamma_1 \, \,\,\,\,
P^T = -k_{33} k_{21}  \gamma_3 \gamma'_a \gamma_2 \gamma_1 \,
\ee
Taking into account the anticommutation rules for Majorana operators, we know that the expectation value of $ \gamma_3 {\tilde\gamma}(\Omegas)  \gamma_2 \gamma_1$ in the final state should be equal to the expectation value of  $- \gamma_2 \gamma_a \gamma_3 \gamma_1$ in the initial state. This tells us that
 for $\Omegas=\Omegas_T$, being the parameters at the final state,  the operator $\gamma'_a $ is equal to $s_{\bar{2}  \bar{3}} {\tilde\gamma}(\Omegas_T)$, with   $ s_{\bar{2} \bar{3}} = -k_{22} k_{31}
 k_{33} k_{21} $.  Furthermore, if $A$ is any product of Majorana operators far from the junction, we have
 $\sex {A \gamma'_a}_T = s_{\bar{2} \bar{3}} \sex{A \gamma_a}_0 $, where the subscript indicates the time at which the expectation value is taken.

 As an example of the above,  consider the case where $(\phi_1, \phi_2,\phi_3)  = (\pi/6, 5 \pi/6, 3\pi/2)$. Then $k_{22 }= -1, k_{31}= 1, k_{33} = 1, k_{21} =1$.  In this case we have no sign change: in the final state we have $\gamma'_a$ being equal to the continuously evolving ${\tilde\gamma}$ at time $T$.  On the other hand, if we rotate the system by $- \pi/3$, so that  $(\phi_1, \phi_2,\phi_3)  = (3 \pi/2,\pi/6, 5 \pi/6)$, we find
  $k_{22 }= 1, k_{31}= -1, k_{33} = -1, k_{21} =1,$ so  in the final state we will have $\gamma'_a=   -{\tilde\gamma}(\Omegas_T)$.

 These results can be readily generalized to the case of arbitrary angles between the wires. We find that the factor of -1 occurs if and only if the  negative $x$-axis lies within the area between two lines of orientation  $\phi_3$  and $\phi_2 $, with the restriction that the opening not include the direction $\phi_1- \pi$.

The results derived above can be stated in another way.  For both cases considered in Fig.  \ref{fg:BasicY}, the necessity for including a minus sign at the end of the manipulation can be determined by defining  a continuous path for change in the orientation  angle $\alpha_a$ between the original orientation in wire 2 and the final orientation in wire 3. If this path crosses  the cut at angle $\pi$ mod $2 \pi$, we must multiply the Majorana operator by a minus sign.
 In the case where one incoming wire is occupied far from the junction (upper panels of Fig.  \ref{fg:BasicY} ), we must  define the   path between the original and final angles  by requiring that $\alpha $ is never equal to the wire orientation angle $\phi_2$.  If we think of the continuous path as a set of Majorana orientations which would occur if one continuously bent the end of wire 2 until it aligned with wire 3,  the condition effectively says that the wire would not double back on itself.  In the case of three full wires, illustrated in the lower panels of the figure, the path should be chosen by requiring that $\alpha$ {\em must} pass through the orientation angle $\phi_1$ of the third wire at the junction.

If  the $Y$ junction is not oriented parallel to the x-y plane, or if the orientation of $\hat{\bf b}$ is not parallel to $\hat{\bf z}$,  the factors $k_{ij}$ in the parity operators will no longer be given simply by Eq.~(\ref{kij}). Nevertheless, the relations
 $s_{\bar{2}3} = k_{22} k_{23}$  and  $ s_{\bar{2} \bar{3}} = -k_{22} k_{31}
 k_{33} k_{21} $ still hold, as they followed directly from conservation of overall parity.


\section{Majorana exchange}
\label{se:MajornaExchange}
The process of exchanging two Majoranas can now be performed by compounding  elementary processes of the type described above, where a  Majorana is moved through a $Y$ junction from one wire to another.

\begin{figure}[h] \begin{center}
 \includegraphics*[width=.9\columnwidth]{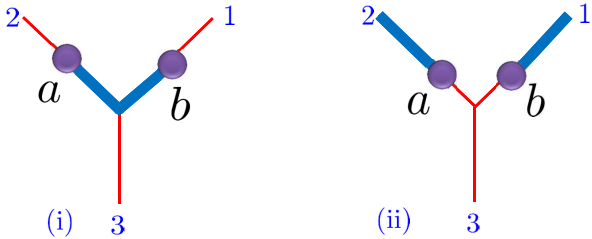}
  \end{center} \vspace{-.5cm} \caption{\label{fg:Mex} Two initial states for a
Majorana exchange process. (i). Two Majoranas on the same TS segment.  (ii).  Majoranas on different TS segments.
} \end{figure}

 An exchange process has two distinct flavors:
either the exchange is between two Majorana states at the edges of the
same topological segment, or between states belonging to different
segments. A configurations of each type is illustrated in Fig.~\ref{fg:Mex}.

When the Majorana states belong to the same topological segment, an
exchange can be described as the following 'three-point turn' sequence:
\be
\label{exchangeab}
^{a}Y^{b}\stackrel{s_{23}}{\longrightarrow} {Y_{a}^{b}}\stackrel{s_{12}}{\longrightarrow}{^{b}Y_{a}}
\stackrel{s_{31}}{\rightarrow} {^{b}Y^{a}}
\ee
The sign change of $\gamma_a$ as it shifts from branch 2 to branch 1 via
branch 3 is given by the product of  two signs:
\be
\gamma_a\rightarrow s_a \gamma_a \, , \,\,\,\,
s_a = s_{23}s_{31}
\ee
For $\gamma_b$, the result is
\be
\gamma_b\rightarrow s_b \gamma_b \, , \,\,\,\,
s_b = s_{12}.
\ee
The arrow notation here implies that in computing the expectation value of a product of Majorana operators which includes $\gamma_a, \gamma_b$ or both, the expectation value after the exchange will be multiplied by a sign equal to $s_a, s_b$ or $s_a s_b$.  Note that in our  labeling scheme, the Majorana operators retain their original subscripts, even though their positions have been interchanged. We have now dropped the prime symbol used in the previous section to designate Majorana states in the final configuration.

The values of the signs $s_{ij}$ can be obtained from the results of the previous section, using a permutation of the labels 1,2,3 where necessary.  Each step in (\ref{exchangeab}) is composed of two elementary parts.  For example, in the first step, Majorana $a$ is pulled back through the junction from wire 2 to wire 1; then it is pushed through the junction onto wire 3.  Thus we have  $s_{23} = s_{2 \bar{1}}  s_{\bar{1} 3}$.
We may treat  this process as a continuous rotation of the Majorana  from the initial direction $\alpha_a = \phi_2 - \pi$ to the final direction $\phi_3 - \pi$. The rotation must be done in the counterclockwise direction, so that $\alpha_a$ is never equal to $\phi_1-\pi$. The sign $s_{23}$ will be negative if $\alpha$ crosses the negative $x$-axis in the process.   Similarly, $s_{12}$ will be negative if $\alpha_b$ crosses the negative $x$-axis when it is rotated in a counterclockwise sense from its initial direction $\phi_1 - \pi$ to $\phi_2-\pi$. The sign $s_{31}$ will be negative if $\alpha_a$ crosses the negative $x$-axis as it is rotated from $\phi_3 - \pi$ to $\phi_1-\pi$.  If we choose the wire directions as suggested by the letter $Y$, say with $(\phi_1, \phi_2, \phi_3 )= ( \pi/6 , 5 \pi/6, 3 \pi / 2)$, we find that  $s_{23} =s_{12} = 1, s_{31} = -1$, so $s_a= -1, s_b=1$.

Note that if we repeat the entire process twice, thereby restoring $a$ and $b$ to their original wires, both Majorana operators will be multiplied by -1. This follows from the fact that $s_a s_b = -1$.  If we consider other orientations of the three wires, the sign $s_a$ and $s_b$ may change, but their product will always be  -1.

In the second case of interest to us, the two Majoranas are on the edges of two
different segments, which is described by the $Y$-configuration
$^{\overline{a}}Y^{\overline{b}}$. [See panel (b) of Fig.~ \ref{fg:Mex}.]   Again, a three-step turn can exchange the two
Majoranas:
\be
\leftexp{\bar a}{Y}^{\bar b}\stackrel{s_{\bar 2 3}}{\longrightarrow} \leftexp{t}{Y}_{a}^{\bar b}\stackrel{s_{\bar 1 \bar 2}}{\longrightarrow}\leftexp{\bar b}{Y}^t_{a}
\stackrel{s_{ 3 \bar 1}}{\longrightarrow} \leftexp{\bar b}{Y}^{\bar a}.
\ee
In the first step, Majorana $a$ is pushed through the junction from wire 2 to wire 3; in the second step Majorana $b$ is pushed  from wire 1 into the junction and pulled  onto wire 2; and finally Majorana $a$ is pulled from wire 3 to wire 1.  The associated sign $s_{\bar{2} 3}$ will be negative if the angle $\alpha_a$ passes through the negative $x$-axis while rotating in a clockwise sense from the initial direction $\phi_2$ to the direction  $\phi_3 - \pi$. The sign $s_{3 \bar{1}}$ will be negative if $\alpha_a$ passes through the negative $x$-axis when rotating in a clockwise sense from $\phi_3 - \pi$ to $\phi_1$.   Finally, $s_{\bar{1} \bar {2}}$ will be negative if $\alpha_b$ passes through the negative $x$-axis on rotating from $\phi_1$ to $\phi_2$.  This rotation must also be taken in a counterclockwise sense, so that the $\alpha_b$ avoids the direction $\phi_2-\pi $.

As in the previous case,  we may define signs $s_a$ and $s_b$ characterizing the change in the Majorana operators after the transformation, with $s_a = s_{\bar {2} 3} s_{3 \bar{1}}$ and $s_b = s_{\bar{1} \bar{2}}$.  Again, we find that the product $s_a s_b$ is always negative.
With the particular  choice of orientations specified above, we have $s_{\bar {2} 3}= s_{3 \bar{1}}=1$ and $s_{\bar{1} \bar{2}}= -1$

As a direct application of the above results, we may consider what happens if we begin with a configuration like that shown in panel (b) of Fig.~\ref{fg:Mex}, and we are given the number parities $P_2$ and $P_1$ of two occupied segments in the initial state. If we perform two clockwise  Majorana exchanges of the type described above, we will reach a final state in which the expectation values of $P_2$ and $P_1$ will each have changed sign.  By contrast, if we perform only a single Majorana exchange, $P_1$ and $P_2$ will each have zero expectation value in the final state.  The state will actually  be a coherent superposition, with equal weights, of a state where the parities have changed and one where they are the same as in the initial state.

As pointed out in Ref. \onlinecite{Clarke10}, it is evident from our analysis that the choice of chirality of the underlying p-wave proximity affects the exchange process and determines which Majorana obtains a minus sign upon exchange. In the $N=2$ model, it is our choice for the sign of the spin-orbit coupling that governs this chirality.

\section{Summary}
\label{se:Summary}
The results of the previous analyses may be summarized as follows.    For an arbitrary continuous manipulation of the parameters of the Hamiltonian, consistent with the requirement that the Majorana fermions always remain far apart, we may define a number $q_j$ for each Majorana, which equals $\pm 1$ depending on whether the parameters defining  that Majorana have crossed an even or odd number of cuts where the Majorana wave function changes sign. (This must include any sign change that occurs when the Majorana is passed through a $Y$ junction).  If the changes in the Hamiltonian are adiabatic in the sense that they are slow compared to the scale set by the gap to finite energy excitations, but fast compared to the exponentially small interactions between the separated Majoranas, then the expectation value of a product of any even number of Majorana operators $\gamma_j$ taken after the manipulation will be identical to the expectation value before the manipulation, except for a sign factor $Q$ which is the product of the $q_j$ for the Majoranas involved.

Suppose that at the end of a series of adiabatic manipulations the system is divided into a set of disjoint TS wire segments.  The quantities of immediate physical significance will be the electron number parities in the individual   wire segments.
The operator which measures the parity of a TS segment  may be expressed
as a product  $i\gamma_a \gamma_b$ of the Majorana operators at the two ends of the segment
with an overall sign $k_{ab}$ that depends on parameters such as the
orientations of the wire ends, but does not depend on the past history. Rules for computing the factors $k_{ab}$, as well as the history-dependent Majorana factors $q_j$  have been given above, using a particular convention for the signs of the Majorana operators.   If these factor are known, and if the expectation value  $\sex {i \gamma_j \gamma_k}_0 $  in the initial state is known for the two Majoranas which wind up, respectively,  at the two ends $a$ and $b$ of the wire segment in question, we may calculate the expectation value of the parity operator in the final state as the product of the factors $k_{ab} q_j q_k$ and the  initial expectation value $\sex {i \gamma_j \gamma_k}_0$.

In the simplest case, where the ending Hamiltonian is the
same as the starting Hamiltonian, so that the system contains the same geometric arrangement of TS clusters that it had initially, and if each  Majorana is returned
to the same position ${\bf R}_j$ that it had initially,  the factor
$k$ will be {\em the same in the initial and final states}.  Moreover, if the parity of each wire is known in the initial state, then the initial expectation values  $\sex {i \gamma_j \gamma_k}_0$ may be deduced. The parity of a wire segment in the final state will then be equal to the parity in the initial state multiplied by the history-dependent product $q_j q_k$  for the two Majoranas at the end of the segment.  If the Majoranas at the ends of a wire segment originated on the same segment, but are reversed in position in the final state, the parity in the final state will still be determined by the parity  in the initial state and the factor $q_j q_k$, but there will be an additional factor of $-1$ due to the reordering of the operators.

If the Majorana operators at the end of a wire segment in the final state originated on two different wire segments in the initial state,  the parity in the final state will not be determined by the parities of individual wires in the initial state.  If the initial state had a definite parity on each wire segment,  then the expectation value for the parity  of the wire segment in question will be zero in the final state.  However, the expectation value of the product of the parities of several such wire segments  may still be non-zero  (and equal to $\pm1$), if the Majoranas at their ends, collectively,  are  a permutation of the Majoranas from an equal number of wire segments in the initial state.

The analyses we have presented can be applied to the case of a three dimensional network of wires, as well as to a two-dimensional network.   In all cases, the result of a set of adiabatic manipulations will be a topological invariant of the path in parameter space between the initial and final states. If the Majorana manipulations are performed on a planar network of wires and junctions, with $\hat{\bf e}$ perpendicular to the plane, with $\hat{\bf b}$ parallel to $\hat{\bf e}$ and with the phase of $\Delta$ the same for the entire system, and if the configuration of TS wires is the same in the initial and final states,  then the result for any physical quantity will be the same as one would have obtained  by analysis of the braiding of the positions  $\bf{R}_j$. These results would be the same as for the braiding of $e/4$-quasi-particles  in non-Abelian $\nu=5/2$ fractional quantum Hall state \cite{Nayak08}

\section{Acknowledgments}

We thank Charles Marcus and Karsten Flensberg for helpful discussions. The work was funded by  Microsoft grant, NSF grant DMR-0906475, NSF grant DMR-1055522, BSF, Minerva and SPP 1285 of the Deutsche Forschungsgemeinschaft. G.R. acknowledges funding provided by the Institute for Quantum
Information and Matter, an NSF Physics Frontiers Center with support of the Gordon and Betty Moore Foundation. We also acknowledge the hospitality of the Aspen Center for Physics and the Kavli Institute for Theoretical Physics (University of California, Santa Barbara)


\begin{thebibliography}{10}

\bibitem{Kitaev-TQC}
A.~Y. Kitaev, Ann. Phys. {\bf 303},  2   (2003).

\bibitem{Nayak-TQC}
S. {Das Sarma}, M. {Freedman}, and C. {Nayak}, Phys. Rev. Lett. {\bf 94},
  166802  (2005).

\bibitem{MooreRead}
G. Moore and N. Read, Nucl. Phys. B {\bf 360},  362   (1991).

\bibitem{ReadGreen}
N. Read and D. Green, Phys. Rev. B {\bf 61},  10267  (2000).

\bibitem{Volovik-pwave}
G. Volovik, JETP Lett. {\bf 70},  609  (1999).

\bibitem{Beenakker-ES}
M.~Wimmer, A.~R.~Akhmerov, M.~V.~Medvedyeva, J.~Tworzydlo, and
  C.~W.~J.~Beenakker, Phys. Rev. Lett. {\bf 105}, 046803 (2010).

\bibitem{FuKane3d}
L. Fu and C.~L. Kane, Phys. Rev. Lett. {\bf 100},  096407  (2008).

\bibitem{Sau2DEG}
J.~D. Sau, R.~M. Lutchyn, S. Tewari, and S. Das~Sarma, Phys. Rev. Lett. {\bf
  104},  040502  (2010).

\bibitem{Alicea2DEG}
J. Alicea, Phys. Rev. B {\bf 81},  125318  (2010).

\bibitem{ShtengelIF}
P. Bonderson, A. Kitaev, and K. Shtengel, Phys. Rev. Lett. {\bf 96},  016803
  (2006).

\bibitem{SternIF}
A. Stern and B.~I. Halperin, Phys. Rev. Lett. {\bf 96},  016802  (2006).

\bibitem{Kang11}
{H. Choi, P. Jiang, M.~D. Godfrey, W. Kang, S.~H. Simon, L.~N. Pfeiffer, K.~W.
  West, and K.~W. Baldwin}, New J. Phys. {\bf 13},  055007  (2011).

\bibitem{Kang11a}
{Sanghun An, P. Jiang, H. Choi, W. Kang, S. H. Simon, L. N. Pfeiffer, K. W.
  West and K. W. Baldwin}, ArXiv e-prints 1112.3400  (2011).

\bibitem{Willett}
R.~L. Willett, L.~N. Pfeiffer, and K.~W. West, Phys. Rev. B {\bf 82},  205301
  (2010).

\bibitem{Kitaev01}
A.~Y. Kitaev, Physics-Uspekhi {\bf 44},  131  (2001).

\bibitem{Oreg10}
Y. Oreg, G. Refael, and F. von Oppen, Phys. Rev. Lett. {\bf 105},  177002
  (2010).

\bibitem{Lutchyn10a}
R.~M. Lutchyn, J.~D. Sau, and S. Das~Sarma, Phys. Rev. Lett. {\bf 105},  077001
   (2010).

\bibitem{Hassler10}
F. Hassler, A.~R. Akhmerov, C.-Y. Hou, and C.~W.~J. Beenakker, New J. Phys.
  {\bf 12},  125002  (2010).

\bibitem{Potter10}
A.~C. Potter and P.~A. Lee, Phys. Rev. Lett. {\bf 105},  227003  (2010).

\bibitem{FuKane-edge}
L. Fu and C.~L. Kane, Phys. Rev. Lett. {\bf 102},  216403  (2009).

\bibitem{Fu09}
L. Fu and C.~L. Kane, Phys. Rev. B {\bf 79},  161408  (2009).

\bibitem{Ivanov}
D.~A. Ivanov, Phys. Rev. Lett. {\bf 86},  268  (2001).

\bibitem{SternvonOppen}
A. Stern, F. von Oppen, and E. Mariani, Phys. Rev. B {\bf 70},  205338  (2004).

\bibitem{Alicea10}
{J. Alicea, Y. Oreg, G. Refael, F. von Oppen, and M.~P.~A. Fisher}, Nature
  Phys. {\bf 7},  412  (2011).

\bibitem{Clarke10}
D.~J. Clarke, J.~D. Sau, and S. Tewari, Phys. Rev. B {\bf 84},  035120  (2011).

\bibitem{Sau11}
J.~D. Sau, D.~J. Clarke, and S. Tewari, Phys. Rev. B {\bf 84},  094505  (2011).

\bibitem{VanHeck11}
{B.~van Heck, A.~R.~Akhmerov, F.~Hassler, M. Burrello, and C.~W.~J.Beenakker},
  ArXiv e-prints 1111.6001  (2011).

\bibitem{Romito11}
A. {Romito}, J. {Alicea}, G. {Refael}, and F. {von Oppen}, ArXiv e-prints
  1110.6193  (2011).

\bibitem{Teo10}
J.~C.~Y. Teo and C.~L. Kane, Phys. Rev. Lett. {\bf 104},  046401  (2010).

\bibitem{Freedman11a}
{M. Freedman, M. B. Hastings, C. Nayak, X.-L. Qi, K. Walker,} and {Z. Wang},
  Phys. Rev. B {\bf 83},  115132  (2011).

\bibitem{Freedman11}
M. {Freedman}, M.~B. {Hastings}, C. {Nayak}, and X.-L. {Qi}, ArXiv e-prints
  1107.2731  (2011).

\bibitem{Fu08}
L. Fu and C.~L. Kane, Phys. Rev. Lett. {\bf 100},  096407  (2008).

\bibitem{Brouwer11}
P.~W. Brouwer, M. Duckheim, A. Romito, and F. von Oppen, Phys. Rev. B {\bf 84},
   144526  (2011).

\bibitem{Nayak08}
{C. Nayak, S.~H. Simon, A. Stern, M. Freedman, and S. Das Sarma}, Rev. Mod.
  Phys. {\bf 80},  1083  (2008).

\end{thebibliography}

\end{document}